\documentclass[lettersize,journal]{IEEEtran}
\usepackage{amsmath,amsfonts}
\usepackage{algorithm}
\usepackage{array}
\usepackage[caption=false,font=footnotesize]{subfig}
\usepackage{textcomp}
\usepackage{stfloats}
\usepackage{url}
\usepackage{verbatim}
\usepackage{graphicx}
\usepackage{cite}
\usepackage{algpseudocode}
\usepackage{hyperref}
\usepackage{color}
\usepackage{multirow}

\usepackage[labelformat=simple]{subcaption}

\begin{document}

\title{Vibrato Expression Control for Singing Voice Conversion 
with Improving Independent Control}

\author{
Joon-Seung Choi\href{https://orcid.org/0009-0003-5791-1569}{\includegraphics[scale=0.5]{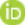}}, 
Dong-Min Byun\href{https://orcid.org/0009-0002-7587-2244}{\includegraphics[scale=0.5]{figure/orcid/orcid_16x16.pdf}}, 
and Seong-Whan Lee\href{https://orcid.org/0000-0002-6249-4996}{\includegraphics[scale=0.5]{figure/orcid/orcid_16x16.pdf}},~\IEEEmembership{Fellow,~IEEE}


\thanks{This work was partly supported by Institute of Information \& Communications Technology Planning \& Evaluation (IITP) under the artificial intelligence graduate school program (Korea University) (No. RS-2019-II190079) and artificial intelligence star fellowship support program to nurture the best talents (IITP-2025-RS-2025-02304828) grant funded by the Korea government (MSIT).). \textit{(Corresponding author: Seong-Whan Lee.)}}
\thanks{J.-S. Choi, D.-M. Byun, and S.-W. Lee are with the Department of Artificial Intelligence, Korea University, Seoul 02841, South Korea (e-mail: js\_choi@korea.ac.kr; dm\_byun@korea.ac.kr; sw.lee@korea.ac.kr)}}



\markboth{Journal of \LaTeX\ Class Files,~Vol.~14, No.~8, August~2021}%
{Shell \MakeLowercase{\textit{et al.}}: A Sample Article Using IEEEtran.cls for IEEE Journals}



\maketitle

\begin{abstract}
Singing style is a crucial aspect of a natural and expressive singing voice. Singers utilize singing styles to convey the feeling or emotion of the songs. Several works have been proposed to control singing style for making the more expressive singing voice. Recently, VibE-SVC successfully controls vibrato by predicting high-frequency F0 contour. In this paper, we introduce a singing voice conversion framework, called VibE-SVC2, to improve singing style conversion performance and controllability. The model offers control over two types of singing styles: a pitch style and a timbre style. For the pitch style, to resolve the pitch-energy entanglement issue that is unresolved in our previous work, we introduce a novel Energy Style Converter to address remaining style information in the energy contour.
In addition, we propose a Zero-shot Pitch Style Converter, which mimics the pitch style of reference audio. To expand the controllability of the model, we propose vibrato rate scaling that is an independent control of vibrato extent, which is unavailable in VibE-SVC.
For the timbre style, we extend the model to handle a variety of phonation styles. However, addressing specific styles such as vocal fry poses a challenge, as conventional F0 extraction often fails due to their inherent subharmonic characteristics, which degrades the conversion quality.
To address this, we propose a novel Subharmonic Correction algorithm 
to refine the F0 contour for more natural timbre conversion.
Through comprehensive objective and subjective evaluations, we demonstrate that VibE-SVC2 provides fine-grained, independent control over two types of singing styles, outperforming existing methods.
\end{abstract}

\begin{IEEEkeywords}
Singing voice conversion, Singing style conversion, Vibrato rate scaling, Timbre style conversion
\end{IEEEkeywords}

\section{Introduction}
Voice conversion (VC) is a technique that preserves the linguistic information of the source audio while converting speaker timbre, prosody\cite{prosovc, zhao2025prosody}, emotion \cite{pavits,durflex}, and accent \cite{wang2021accent,jin2023voice,macst}. Singing voice conversion (SVC) often builds upon VC, preserving not only the lyrics but also the melody. 
Various generative models for singing voice have emerged, including variational autoencoder (VAE) models \cite{ning2023vits,zhou2023vits,free-svc}, diffusion models \cite{diffsvc,ldm-svc,comosvc,lcm-svc,midivoice,midivoice2}, and flow-matching models \cite{dafmsvc, serenade}. 
Beyond these foundational models, SVC research has advanced to tackle challenges such as timbre leakage \cite{syki-svc,neuco-svc}, robustness \cite{robust-svc, knn-svc}, model efficiency \cite{lhq-svc}, and expressiveness \cite{esvc, serenade, vevo2}.

Among the challenges of SVC, enhancing expressiveness has been actively researched. Improving expressiveness requires the control of various expressive attributes, such as singing style and emotion. The growing importance of Singing Style Conversion (SSC) is highlighted by the recent SVCC 2025 \cite{svcc2025}. However, modeling SSC poses a significant challenge due to the dynamic characteristics of singing voices. Specifically, these characteristics encompass exceptionally wide pitch ranges, large pitch fluctuations, and diverse spectral timbres. Since expressive singing involves simultaneous style attributes across pitch, energy, and timbre, it is difficult to manipulate one stylistic attribute without distorting the others. This difficulty arises because these dynamic attributes are intrinsically coupled and strongly synchronized in human vocal production. Due to this inherent complexity, the baseline models adopted in SVCC 2025 still exhibit limitations. Serenade \cite{serenade} omits the conversion of pitch styles, while Vevo1.5 \cite{vevo2} focuses primarily on linguistic content, melody, and style separation rather than disentangling specific timbre styles from global timbre. These approaches focus only on partial separation in various singing styles. To overcome these limitations directly, we categorize these dynamic attributes into pitch- and timbre-related components, focusing on the specific acoustic couplings that define them.

Controlling pitch-related attributes requires disentangling these complex couplings, a challenge most evident in vibrato. As the most typical singing style, vibrato is acoustically characterized not only by pitch fluctuation, but also by a highly synchronized and periodic oscillation in loudness \cite{psychology_music}. These periodic oscillations are fundamentally defined by two acoustic parameters : the rate and extent. Consequently, providing fine-grained, independent control over these individual attributes is a highly non-trivial task.

Various attempts have been made within Singing Voice Synthesis (SVS) frameworks to address this challenge by predicting exact parameter values or using the first-order difference of the F0 ($\Delta$F0) contour \cite{expressivesing, vibextsvs}. Despite these efforts, these approaches fall short of achieving disentanglement. For instance, one method \cite{vibextsvs} relying on intermediate features correlated with vibrato extent forces the model to learn entangled representations, failing to structurally isolate the vibrato. On the other hand, another model \cite{expressivesing} directly conditioned on precise vibrato parameter values \cite{expressivesing} exhibits sensitivity to the prediction errors.
Consequently, even slight deviations in parameter prediction easily lead to unstable or unnatural generation.

To address this, VibE-SVC \cite{vibesvc} introduced a pitch style converter that effectively separated and converted vibrato characteristics by extracting high-frequency F0 contours via the discrete wavelet transform (DWT). The pitch style converter predicts the high-frequency F0 contour and enables vibrato extent control. Despite this advancement, VibE-SVC had four key limitations in achieving comprehensive and generalized style control.

First, our experiments confirmed that the periodic amplitude modulation of vibrato remains entangled within the energy contour. Since pitch and energy are inherently coupled \cite{psychology_music}, energy fluctuations are highly synchronized with F0 variations. While some recent SVS studies \cite{expressivesing, ryu2025controllable} adopt energy predictors, they treat energy holistically and overlook this periodic modulation. To resolve this pitch-energy entanglement, we introduce a novel Energy Style Converter that can control these vibrato characteristics within the energy contour by leveraging this periodicity.

Second, VibE-SVC was limited to vibrato extent scaling. Building upon this capability, we propose a temporal scaling mechanism for independent vibrato rate control. By applying temporal scaling to the low-frequency F0 contour during inference, our model enables fine-grained control over both vibrato extent and rate.

Third, the previous framework was restricted to using predefined style IDs. To overcome this limitation and enhance practical applicability, we introduce a Zero-Shot Style Converter (ZSC), allowing the model to transfer pitch styles in F0 and energy contour directly from an unseen reference audio sample without explicit style labels.

Finally, VibE-SVC was solely focused on pitch-related styles, leaving timbre-related styles unexplored. In this work, we expand the framework to independently control a diverse range of phonation styles, including breathy, belt, and vocal fry. While incorporating the spectral variations of most timbre styles is relatively straightforward, handling specific techniques like vocal fry poses an inherent challenge for conventional SVC pipelines. As discussed in previous studies \cite{nansy, gci}, where subharmonic characteristics are accounted for to address the inherently ill-defined nature of the F0 in rough phonation, which produces complex subharmonic patterns. During conversion, these patterns lead to severe pitch-jump
artifacts and unstable timbre. To overcome this, we propose a novel subharmonic correction (SHC) algorithm, which leverages the $\Delta$F0 contour to refine the F0 contour.

This paper extends our previous work, VibE-SVC \cite{vibesvc}, by addressing its inherent limitations and expanding its capabilities into a unified framework, \textbf{VibE-SVC2}. The main contributions beyond our previous work are summarized as follows:

$\bullet$ 
We propose a novel Energy Style Converter that effectively eliminates residual style leakage by resolving the inherent pitch-energy entanglement.

$\bullet$ 
We introduce a rate scaling method that enables independent control over both vibrato rate and extent at inference time.

$\bullet$ 
We propose a Subharmonic Correction (SHC) algorithm that mitigates artifacts during timbre conversion by addressing F0 estimation failures in complex phonations like vocal fry.

$\bullet$ 
Through comprehensive objective and subjective evaluations, we demonstrate that VibE-SVC2 provides independent control over both pitch and timbre styles, outperforming existing methods in style accuracy.

Audio samples\footnote{Demo: \url{https://castlechoi.github.io/VibE-SVC2-demo/}} and code\footnote{Code: \url{https://github.com/castlechoi/VibE-SVC2}} are available on our project website. 

\section{Related Works}
\subsection{Evolution of Expressive Singing Voice Modeling}
The development of deep learning has significantly advanced singing voice generation, transitioning the focus from merely producing intelligible lyrics with correct pitches to modeling highly expressive and realistic performances. To capture the dynamic nuances of human singing, recent frameworks emphasize the independent control of acoustic dimensions such as melody, timbre, and energy. In comprehensive analyses of vocal repertoires, expressive singing techniques, which introduce rich expressive nuances to raw acoustic components, are systematically categorized into pitch- and timbre-related attributes \cite{cosian, sintechsvs}. To establish control criteria for our study, we apply this classification scheme to the VocalSet \cite{vocalset} dataset, allowing us to leverage its diverse phonation styles under a standardized categorization. Following this literature, we define ``pitch singing style'' as techniques modulating the fundamental frequency such as vibrato, and ``timbre singing style'' as techniques manipulating the local spectral characteristics such as vocal fry, breathy, and belt. A straight voice is explicitly defined as the baseline absence of these specific techniques \cite{sintechsvs}. As noted in \cite{sintechsvs}, vocal performances frequently exhibit significant temporal overlaps between pitch and timbre techniques, indicating their simultaneous utilization. Therefore, to achieve fine-grained expressive control in singing style conversion, it is essential that these two acoustic attributes be modeled and manipulated independently without mutual interference, allowing for both isolated control and simultaneous application.  

\subsection{Pitch and Melody Control in Singing Synthesis and Conversion}

In SVS, precise melody control has long been an essential requirement for generating highly natural vocal performances. While its importance is well-established, recent studies continue to emphasize the necessity of capturing fine-grained, singer-specific pitch expressiveness, underscoring that achieving such control remains non-trivial. For instance, StylePitcher \cite{stylepitcher} points out that existing pitch curve generators often neglect individual singing styles and fail to generalize across diverse tasks, reaffirming the ongoing need for robust, general-purpose pitch modeling. As observed in CoMelSinger \cite{comelsinger}, directly extending discrete codec-based speech synthesis to SVS is challenging because prompt-based generation frequently suffers from prosody leakage, a phenomenon where pitch information inadvertently becomes entangled with the timbre prompt, thereby compromising controllability.

Translating this strict requirement and zero-shot capability to SVC introduces fundamentally distinct challenges. Unlike SVS, which typically generates pitch from clean and structured symbolic scores, achieving this paradigm in SVC requires extracting and transferring the melody directly from raw waveforms without explicit labels. This reliance on raw waveforms makes it inherently difficult to decouple the F0 contour from the source speaker's intrinsic vocal characteristics. Acknowledging this fundamental complexity, recent unified frameworks such as Vevo2 \cite{vevo2}  note that designing a robust prosody tokenizer and enhancing the melody encoding capacity of content-style tokenizers remain directions for future exploration.

At the core of this precise melody control is the execution of localized, pitch-related singing styles such as vibrato, scoop, bend, and glissando. Earlier approaches attempted to control pitch style by modeling the first-order difference of the F0 contour ($\Delta$F0) \cite{vibextsvs} or the high-frequency F0 contour \cite{vibesvc}. However, these works demonstrated only the controllability of vibrato extent, leaving the vibrato rate uncontrolled. Furthermore, accurately transferring these techniques involves the complex dynamics of both pitch and intensity. Classical acoustic literature \cite{psychology_music} explicitly highlights this strong synchronization, noting that pitch fluctuations are typically accompanied by amplitude variations of 2-3 dB. As conventional conversion methods often fail to explicitly decouple this inherent pitch-energy synchronization when extracting F0 contour, they suffer from severe pitch-energy entanglement. Consequently, residual stylistic information leaks into the target speaker's acoustic representations, degrading the conversion quality. To address this, our framework structurally isolates pitch dynamics from energy fluctuations. This explicit isolation establishes a robust foundation for both fully independent parameter manipulation and zero-shot pitch style conversion, ensuring fine-grained control without distorting the target's natural energy profile.

\subsection{Timbre and Phonation Style Control in Singing Voice}
Timbre and phonation styles are shaped by vocal tract configurations and glottal behaviors, resulting in diverse vocal expressions such as breathy, belt, falsetto, and straight voice. Effectively controlling these attributes requires manipulating the spectral envelope and harmonic structures independently of the underlying melody. Among these diverse phonation types, vocal fry---also called creaky voice---presents a unique acoustic challenge. It is the lowest vocal register, characterized by a low F0 range and subharmonic patterns. According to a previous study \cite{vocalfry_register}, vocal fry is a specific type of creak. A major difference between the two terms lies in the regularity of the F0 contour. Unlike other types of creaky voice that are characterized by irregularity, vocal fry is often periodic \cite{acoustic_creaky}.

Extensive research has explored vocal fry and creaky voice across various applications. While DeepFry \cite{deepfry} and Creapy \cite{creapy} focus primarily on detection tasks
several architectures have been proposed for generation. For instance, modified neural HMM TTS \cite{lameris2023prosody} and NANSY \cite{nansy} address the ill-defined F0 caused by the subharmonic nature of vocal fry, and CreakyVC \cite{creakvc} proposed a voice conversion framework specifically targeting the style transfer of creaky voice. According to these studies, the inherent subharmonic patterns of vocal fry severely degrade the naturalness of timbre conversion. To mitigate this, our goal is to systematically correct the F0 contour prior to the conversion process to achieve highly natural timbre conversion.

\subsection{Singing Style Transfer and Conversion Frameworks}

Efforts to explicitly control expressive singing styles has been extensively explored in SVS. Recent frameworks have achieved fine-grained controllability by directly generating specific stylistic attributes from musical scores or text prompts. For instance, SinTechSVS \cite{sintechsvs} introduced an attention-based local-score module to improve the structural modeling of specific singing techniques. Similarly, TechSinger \cite{techsinger} employed a prompt-based flow-matching generative model, allowing users to precisely define and control vocal techniques. Furthermore, TCSinger \cite{tcsinger} proposed a clustering style encoder to enhance zero-shot style transfer directly from reference audio.

While SVS models demonstrate fine-grained control over dynamic styles, applying this to SVC presents a distinct challenge. Conventional SVC models have primarily focused on global timbre replacement. For example, SoVITS-SVC utilizes self-supervised learning (SSL) features to disentangle linguistic content from a target speaker's global identity. Recent state-of-the-art zero-shot frameworks have further advanced this global capability. NeuCoSVC \cite{neuco-svc} addresses timbre leakage by substituting source SSL features with reference counterparts, enhanced by large-scale pre-training in NeuCoSVC2. Similarly, SeedVC \cite{seedvc} employs a diffusion transformer combined with an external timbre shifter to effectively perturb and transfer the timbre of source audio. However, because these frameworks treat the singing voice as a macroscopic acoustic target, they lack explicit mechanisms to independently manipulate localized dynamic styles, such as vibrato rate or vocal fry phonation.

To bridge this gap, dedicated singing style conversion (SSC) frameworks have emerged. Serenade \cite{serenade} achieves a controllable SSC framework that utilizes cyclic training and audio infilling to learn style information. The model enhances naturalness via post-processing with a SiFiGAN \cite{sifigan} vocoder. Vevo1.5 and Vevo2 \cite{vevo2} present a unified framework for both speech and singing, employing a prosody tokenizer for melody adjustment and a content-style tokenizer to disentangle style from linguistic content.

Despite these advancements, existing frameworks still lack explicit and decoupled mechanisms for the independent and fine-grained manipulation of localized dynamic singing styles. Furthermore, while zero-shot global timbre conversion has seen significant progress, achieving zero-shot pitch style conversion while maintaining independent control over timbre remains an open challenge. In this work, we aim to incorporate both pitch-and timbre-related singing styles into a unified framework, allowing for the strictly independent control of localized dynamic styles without interference.

\begin{figure*}[!t]
\centering
\includegraphics[width=5.7in, height=3in]{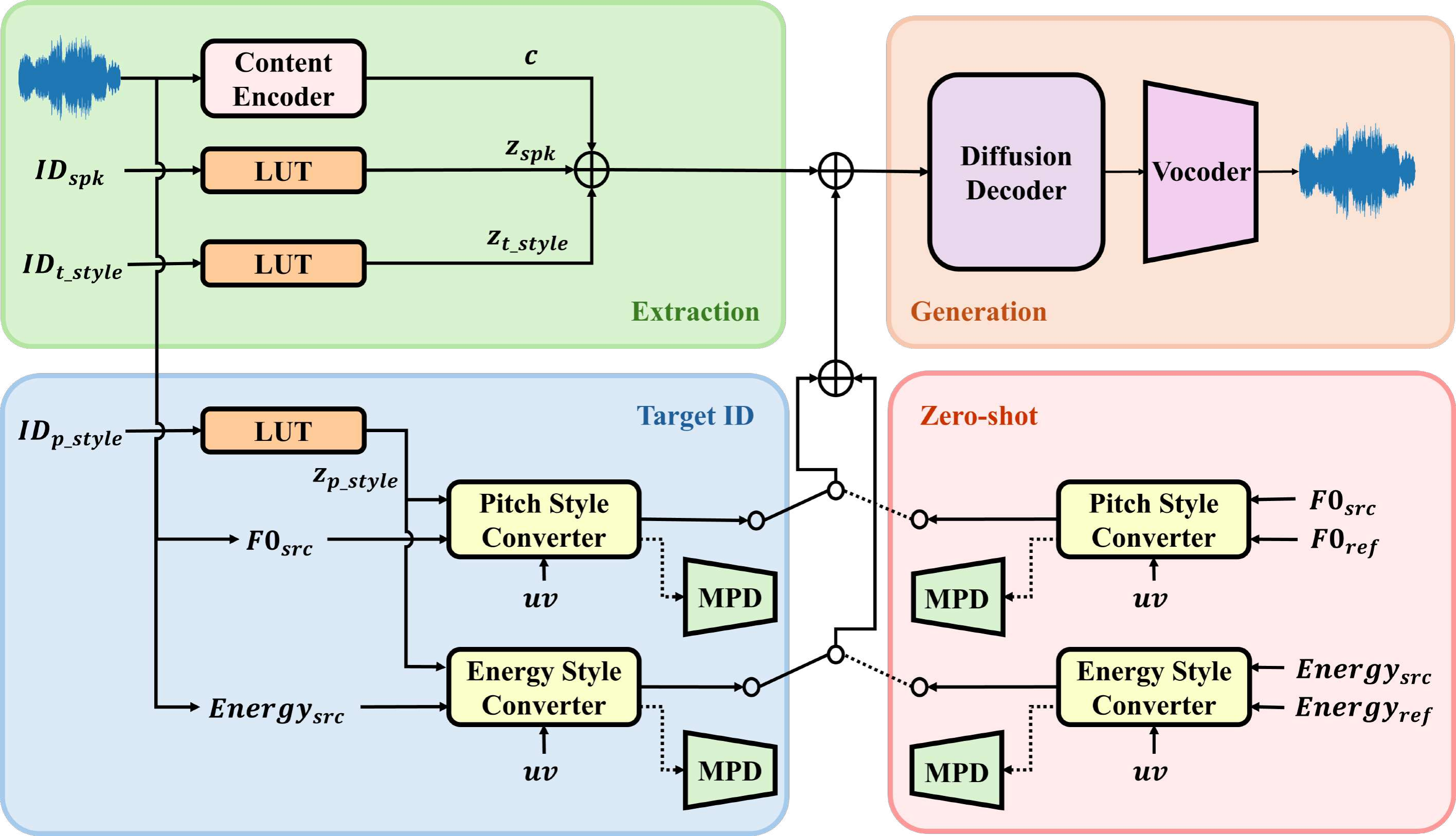}
\caption{The architecture of the VibE-SVC2. ID$_{p\_style}$ and ID$_{t\_style}$ denote the target IDs for the pitch and timbre styles.
}
\label{fig:entire_model}
\vspace{-5px}
\end{figure*}

\section{Methods}

\begin{figure}[!t]
\centering
\subfloat[ID-based Style Converter]{\includegraphics[width=1.6in, height=2.6in]{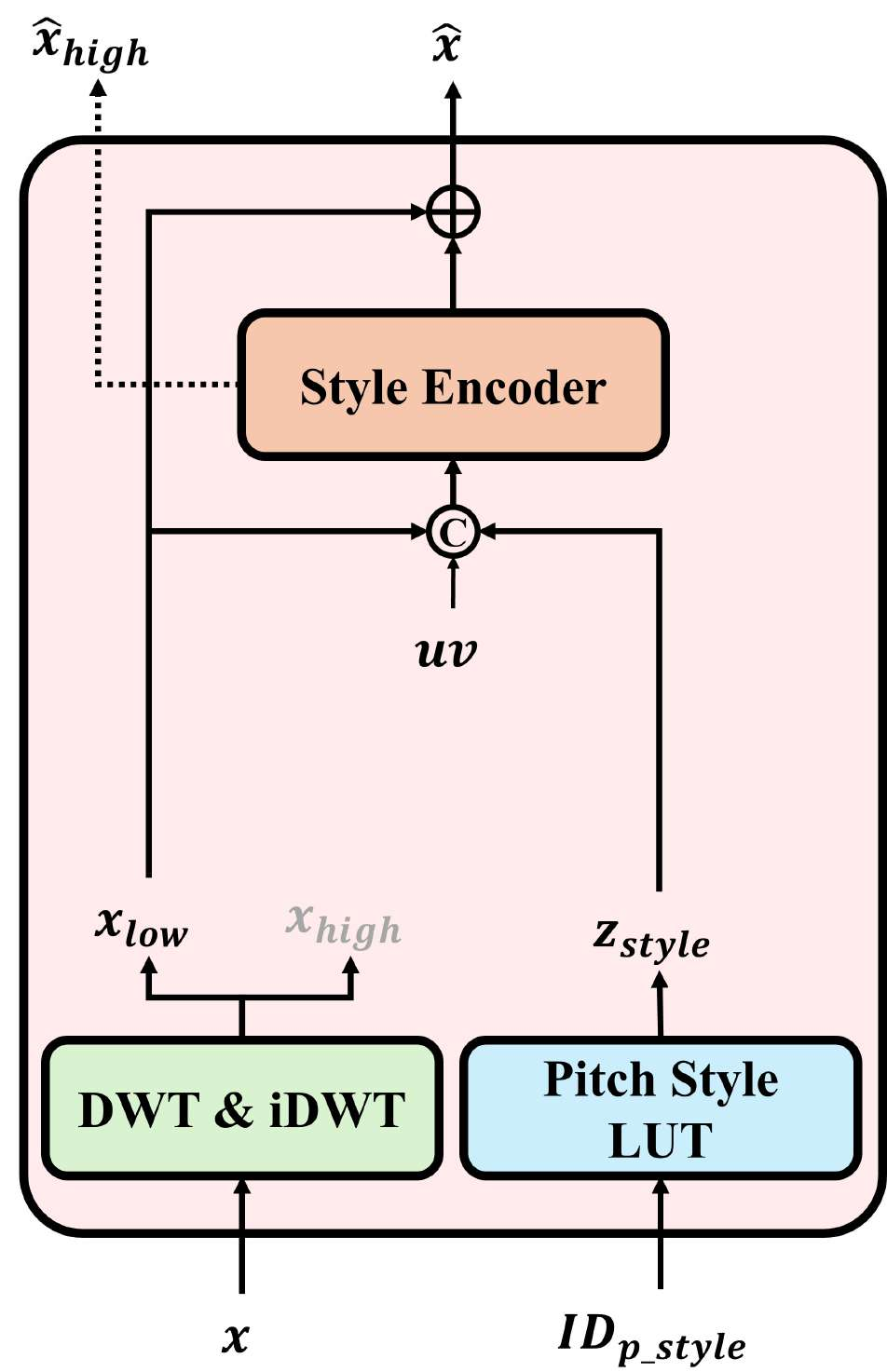}
  \label{fig:ID_pitch_style_converter}}
\hfil
\subfloat[Zero-Shot Style Converter]{\includegraphics[width=1.6in, height=2.6in]{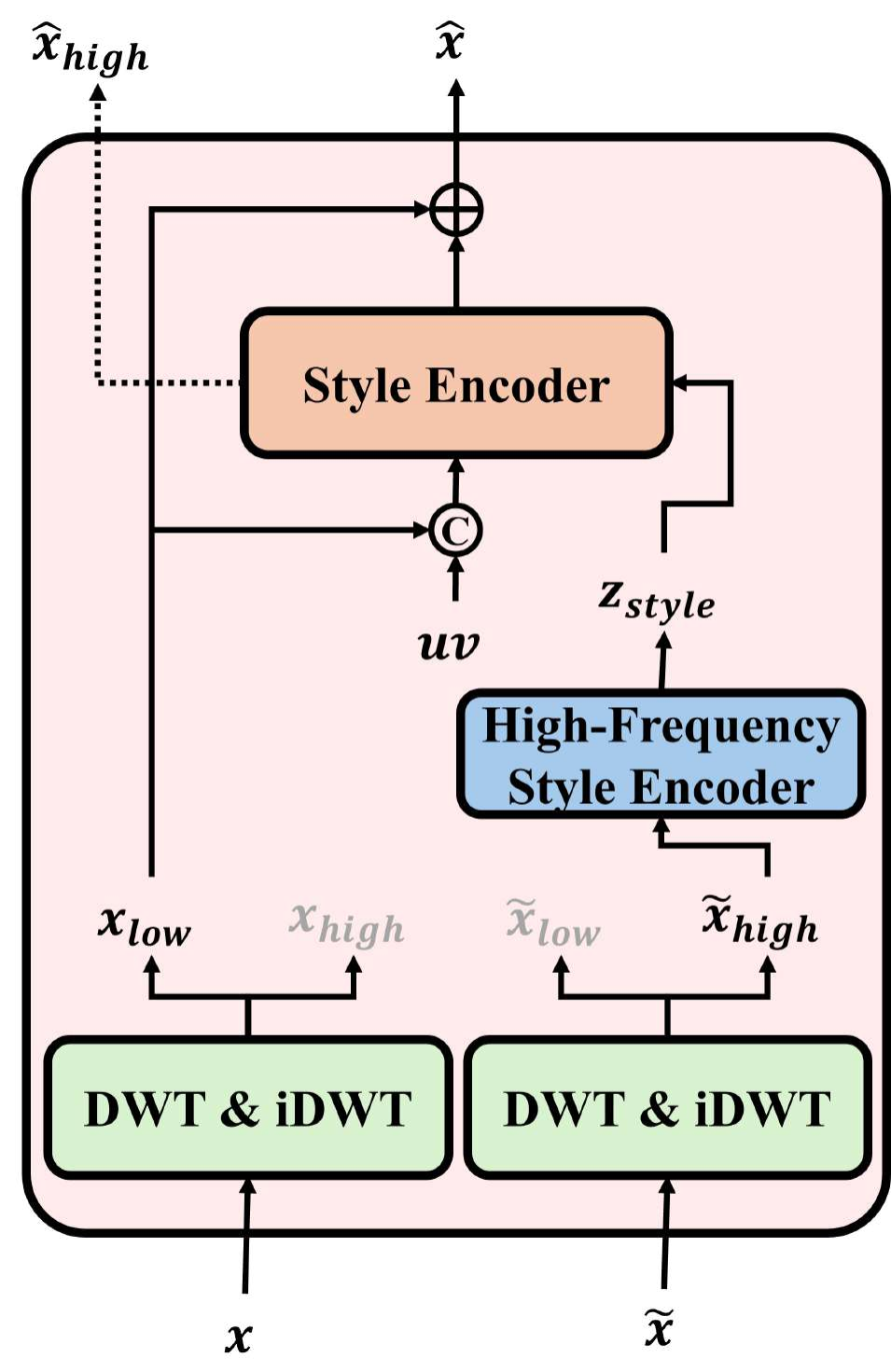}
\label{fig:zs_pitch_style_converter}}

\caption{The architectures of the (a) ID-based Style Converter and the (b) Zero-Shot Style Converter.}

\label{fig:pitch_style_converter}
\vspace{-5px}
\end{figure}

\subsection{Preliminary}
A core principle of our framework is the structural isolation of pitch dynamics from other acoustic attributes. This leads the pitch style converter to be designed to modify the F0 contour. This design ensures that stylistic pitch variations are modeled independently, preventing information leakage into the timbre or energy. Based on this, VibE-SVC \cite{vibesvc} utilizes the DWT and iDWT to disentangle pitch styles from the overall melody. Specifically, DWT decomposes the F0 contour
into an approximation coefficient $A$ and detail coefficients $D$ as follows:
 \begin{align}
    A_j[k] &= \sum_nx[n]\phi_{j,k}[n], \label{equation:dwt_approx} \\
    D_j[k] &= \sum_nx[n]\psi_{j,k}[n],\label{equation:dwt_detail}
\end{align}
where $x$, $j$, $k$, and $n$ denote the source F0 contour, the cutoff level of DWT, the position of wavelet function, and the position of signal, respectively. $\phi$ and $\psi$ denote the scaling function and the wavelet function.
After decomposing the contour into the coefficients, the low- and high-frequency F0 contour are reconstructed from the respective coefficients using iDWT as follows:
\begin{align}
x_{low}[n] &= \sum_kA_{L}[k]\phi_{L,k}[n], \label{equation:iDWT_low} \\
x_{high}[n] &= \sum_{j=1}^L\sum_kD_j[k]\psi_{j,k}[n],\label{equation:iDWT_high}
\end{align}
where $x_{low}$ and $x_{high}$ denote low- and high-frequency F0 contour, respectively.

After separating the F0 contour, the pitch style converter predicts the high-frequency F0 contour from the low-frequency F0 contour and the target style ID.
The pitch style converter is trained with an objective function that consists of a reconstruction loss for the log high-frequency F0 contour, feature matching loss, and adversarial loss.
The reconstruction loss $L_{recon}$ is defined as follows:
\begin{align}
L_{recon}(G) &= MSE(x_{high}, \hat{x}_{high}), \label{equation:recon_loss}
\end{align}
where $G$ denotes the generator and $MSE$ denotes mean square error.
The losses are defined as 
\begin{align}
L(G) &= L_{recon}(G)+\mathcal L_{fm}(G)+\mathcal L_{adv}(G), \label{equation:generator_loss}\\
L(D) &= L_{adv}(D), \label{equation:discriminator_loss}
\end{align}
where $D$ denotes the discriminator.

The work presents a vibrato extent control method by multiplying the high-frequency F0 contour by a constant scaling factor. The formula is as follows:
\begin{align}
x[n] &= x_{low}[n] + \alpha \cdot x_{high}[n], \label{equation:vib_ext_control}
\end{align}
where $\alpha$ denotes the scaling factor for vibrato extent.

\subsection{Energy Style Converter}
\begin{figure}[!t]
\centering
\subfloat[GT]{\includegraphics[width=3.1in, height = 0.7in]{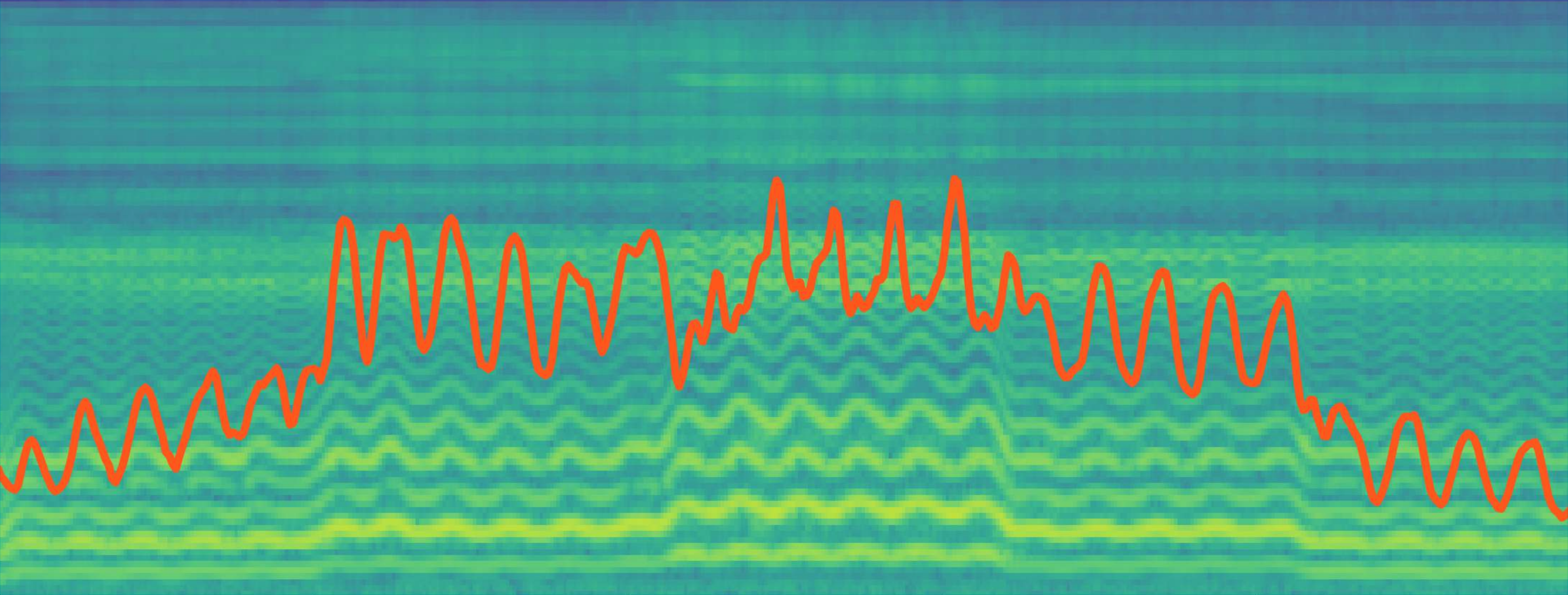}}
\label{fig:energy_gt}

\subfloat[VibE-SVC]{\includegraphics[width=3.1in, height = 0.7in]{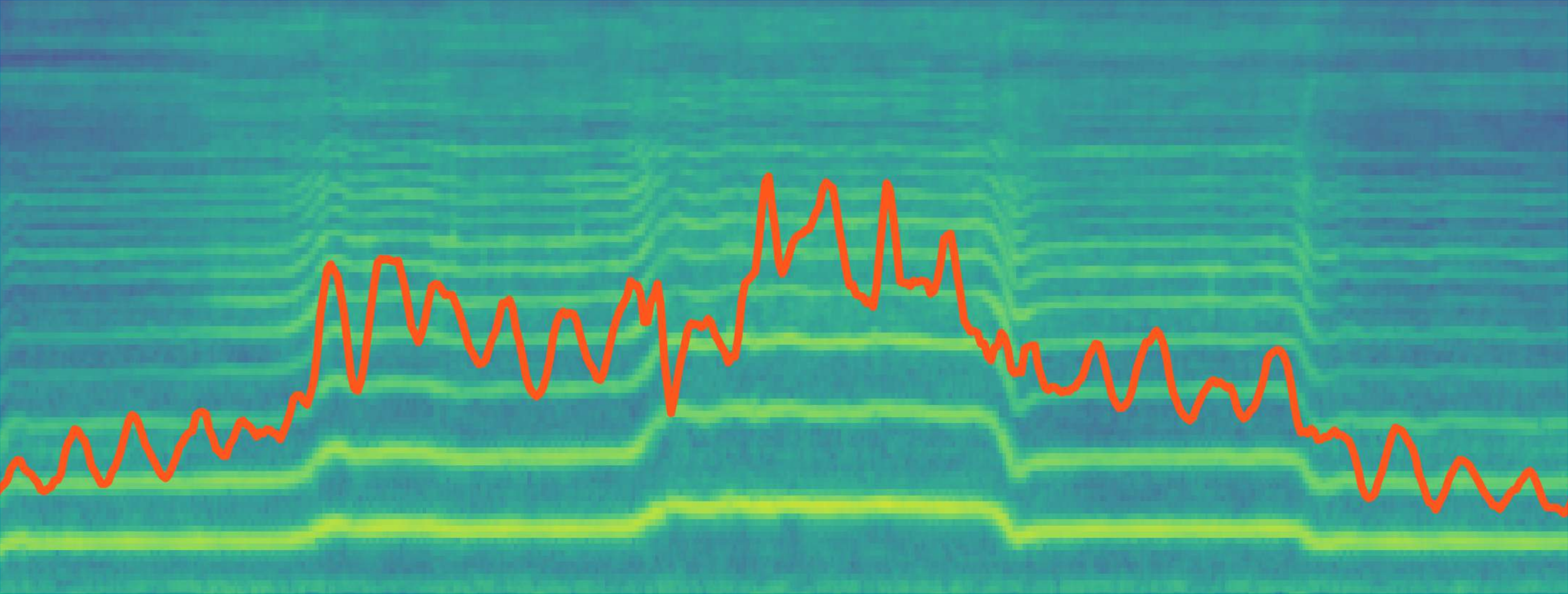}
\label{fig:energy_vibesvc}}

\subfloat[VibE-SVC2]{\includegraphics[width=3.1in, height = 0.7in]{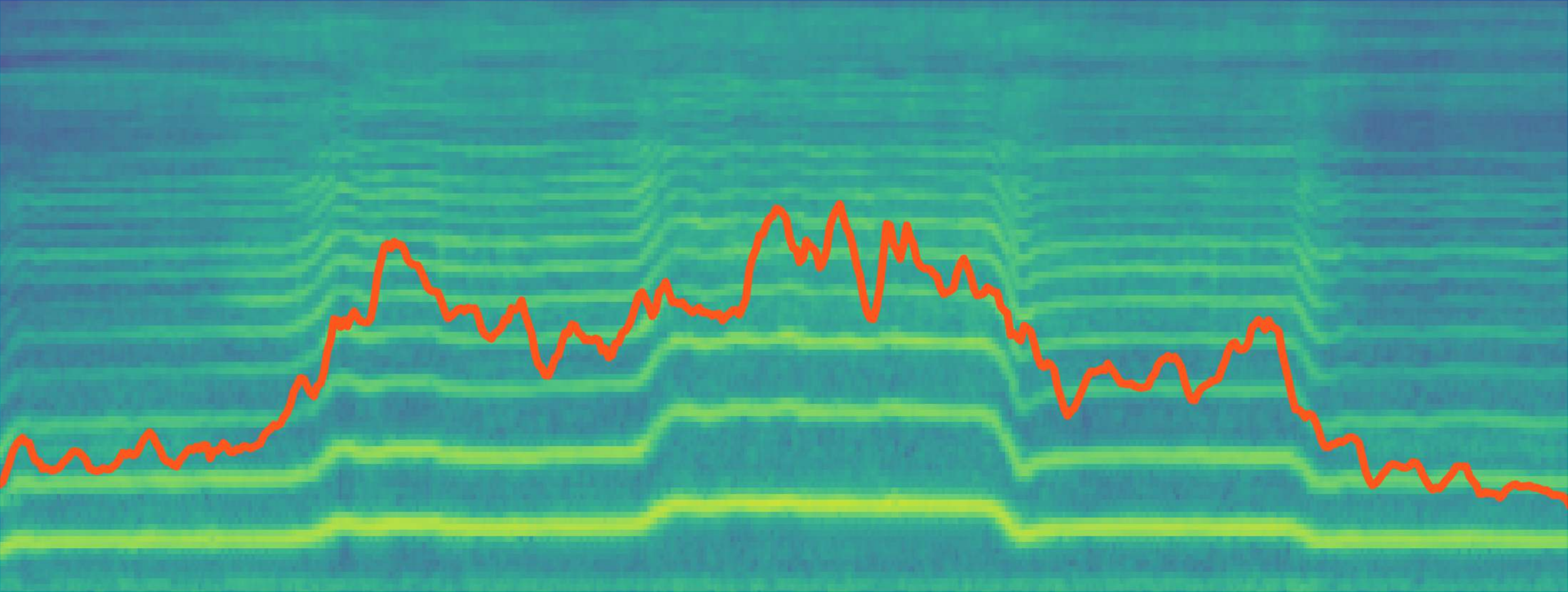}
 \label{dif:energy_vibesvc2}}

\caption{Mel-spectrograms and energy contours during Vibrato$\rightarrow{}$Straight conversion. \label{fig:energy_analysus}}
\vspace{-5pt}
\end{figure}

To improve the performance of pitch style conversion, we introduce an energy style converter by considering the remaining pitch style information within the energy contour. 
As illustrated in Fig. \ref{fig:entire_model}, the energy style converter converts the energy contour into the target style to being processed by the pretrained SVC model. The architecture of the energy style converter is similar to the pitch style converter, as illustrated in Fig. \ref{fig:pitch_style_converter}. One difference is that the converter does not apply a log transformation to the energy contour. It takes the low-frequency energy contour and a target style ID as input. Then, it predicts the corresponding high-frequency energy contour. The predicted high-frequency energy contour is taken from the pretrained SVC model after being added with the low-frequency energy contour. This pipeline leads the decoder to generate the style converted mel-spectrogram based on the predicted energy contour, as shown in Fig. \ref{fig:energy_analysus}. 

\subsection{Zero-Shot Pitch Style Conversion}
We further propose a zero-shot pitch style conversion method, enabling the model to transfer a pitch style from a reference audio.
As shown in Fig. \ref{fig:pitch_style_converter}, this pitch style converter is directly conditioned on the high-frequency F0 contour of the reference audio instead of a target style ID. The high-freq F0 style encoder generates $z_{style}$ from the reference high-frequency F0 contour. Then, $z_{style}$ is injected into every layer using the style adaptive layer normalization (SALN) of the style encoder, as illustrated in Fig. \ref{fig:zs_pitch_style_converter}.
During training, the pitch style converter and energy style converter are trained to reconstruct each high-frequency contour of the source audio as the reference. During inference, the high-frequency contours from a target audio are used to convert the pitch style to that of the reference audio.

\begin{figure}[!t]
\centering
\subfloat[Rate scaling ($\beta > 1$)]{\includegraphics[width=1.6in, height=2.9in]{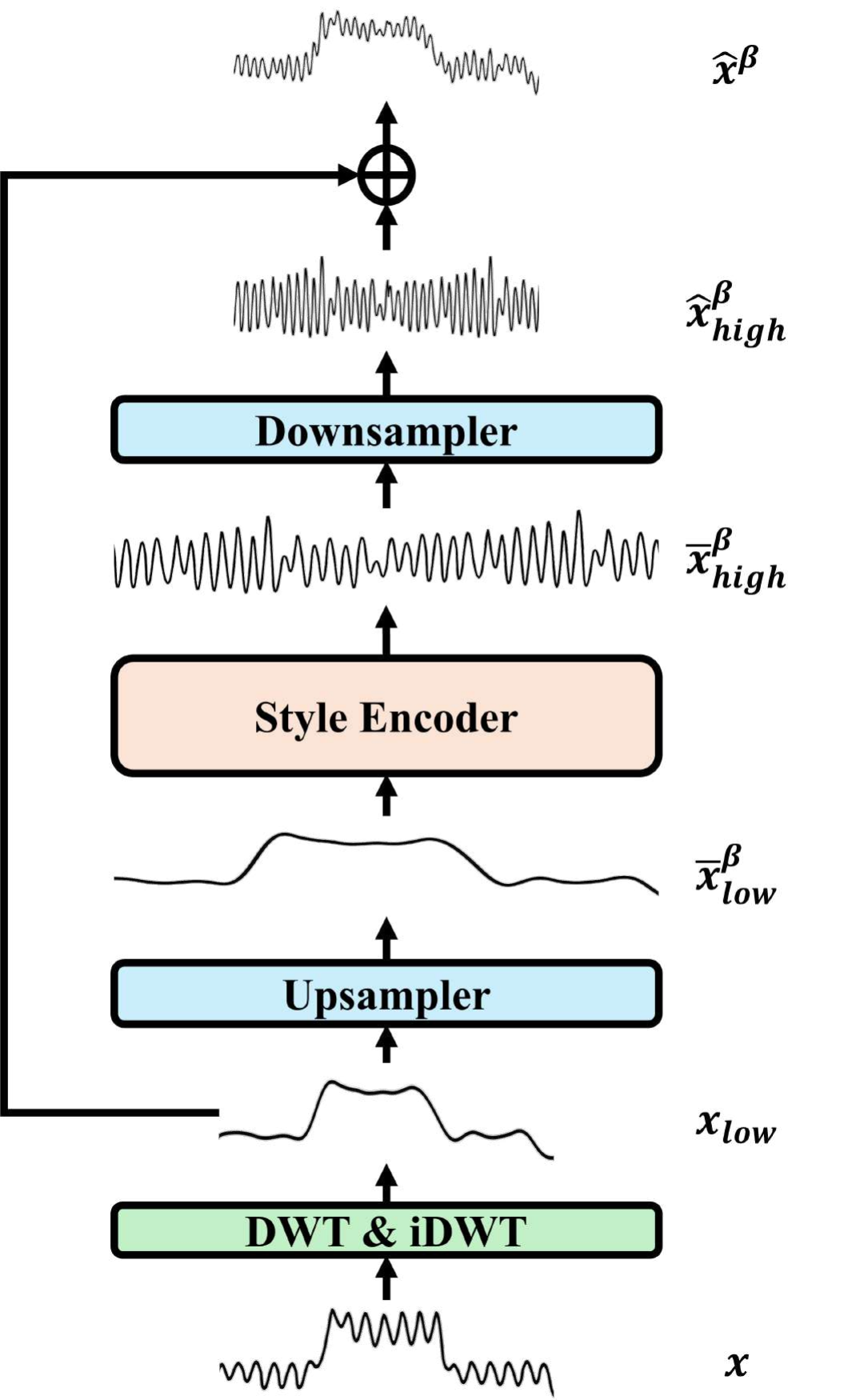}
\label{fig:rate_upsample}}
\hfil
\subfloat[Rate scaling ($\beta < 1$)]{\includegraphics[width=1.6in, height=2.9in]{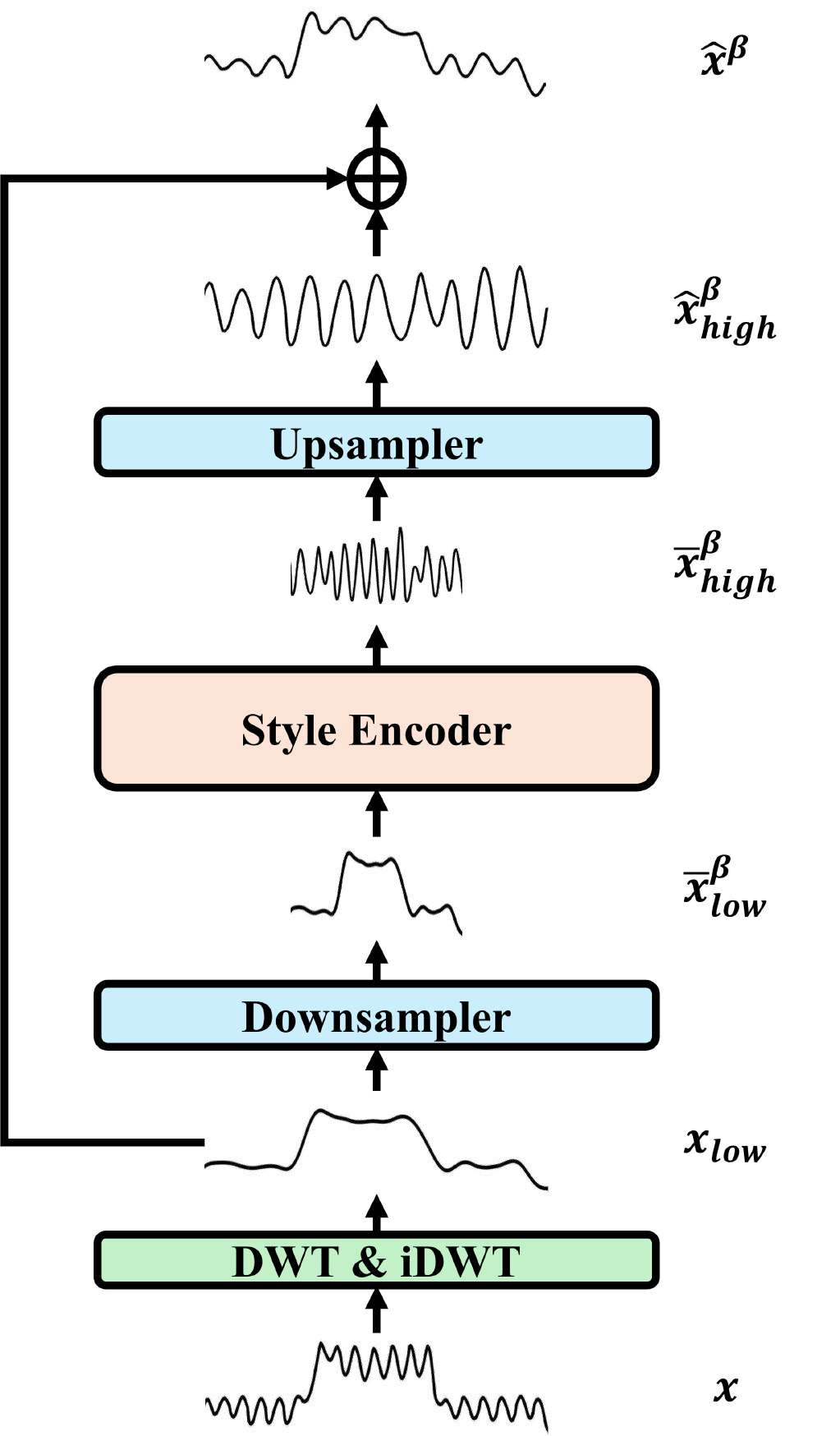}
\label{fig:rate_downsample}}
\caption{
The proposed pipeline for controlling the vibrato rate. (a) shows the process for increasing the vibrato rate and (b) shows the process for decreasing it.}
\label{fig_rate_control}

\end{figure}

\subsection{Vibrato Rate Control}
We propose a vibrato rate control method to provide independent and fine-grained control between vibrato rate and extent. The method is implemented by manipulating the temporal scale of the low-frequency F0 contour that serves as input to the pitch style encoder. As illustrated in Fig. \ref{fig:rate_upsample}, we stretch the low-frequency F0 contour along the time dimension to increase vibrato rate. The encoder generates a corresponding high-frequency F0 contour, which is then compressed back to the original time length. Increasing vibrato rate is formulated as follows:
\begin{align}
{\bar x}_{low}^\beta &= Upsampler\left(x_{low},\beta\right),\\
{\bar x}_{high}^\beta &=StyleEncoder({\bar x}_{low}^\beta,uv,s), \label{equation:rate_control_high} \\
{\hat x}_{high}^\beta &= Downsampler({\bar x}_{high}^\beta, \beta),\label{equation:rate_control_recon}
\end{align}
where $\beta$, $uv$, $s$, and ${\hat x}$ denote a scaling factor of vibrato rate, an unvoiced flag vector, a style embedding, and the predicted output, respectively. 
Conversely, decreasing vibrato rate is implemented by changing the order between $Upsampler$ and $Downsampler$, as illustrated in Fig. \ref{fig:rate_downsample}.
In addition, based on this approach, the vibrato rate and the vibrato extent can be independently controlled at inference time using the following equation :
\begin{align}
\hat x^{\beta}[n] &= x_{low}[n] + \alpha \cdot{\ \hat x}_{high}^\beta[n], \label{equation:recon_raw_f0}
\end{align}
where $\alpha$ denotes the scaling factor for vibrato extent, as previously defined in Eq. \ref{equation:vib_ext_control}.

\begin{figure}[!t]
\centering
\subfloat[Original \& Corrected F0]{\includegraphics[width=3.3in]{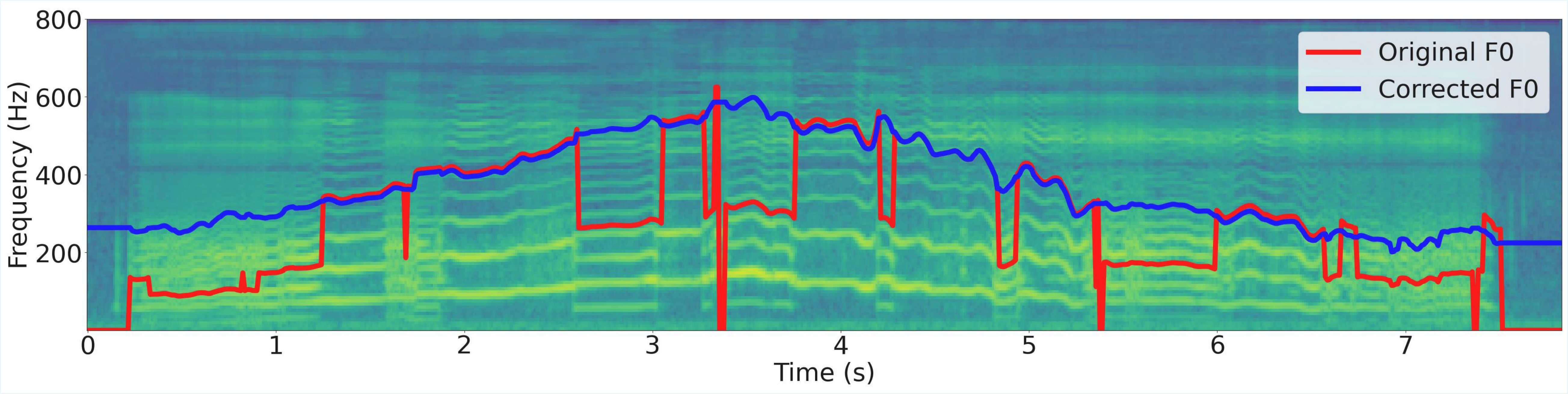}
\label{fig:shc_mel}}
\hfil
\subfloat[Original $\Delta$F0]{\includegraphics[width=3.3in]{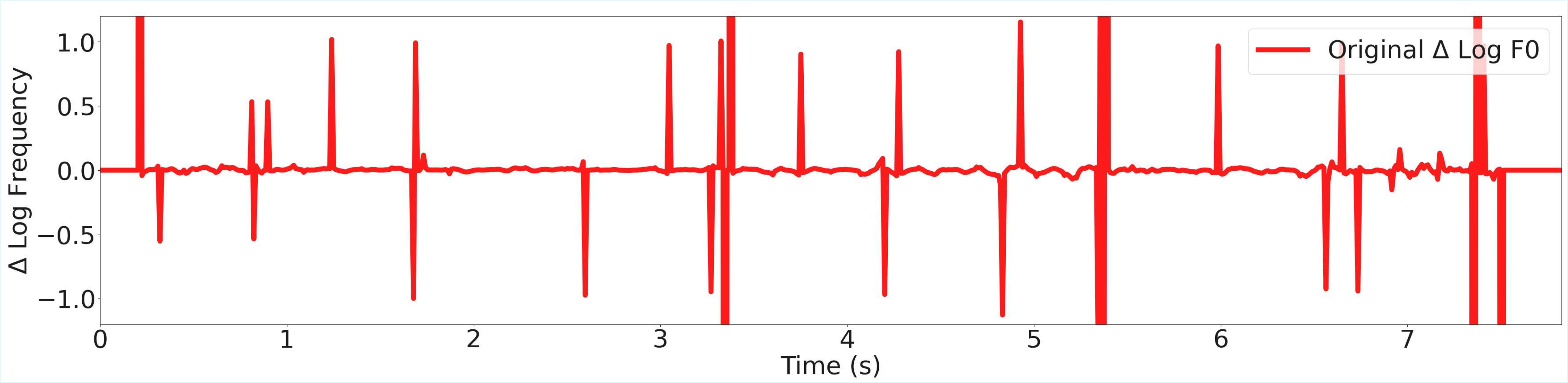}
\label{fig:shc_original}}
\hfil
\subfloat[Corrected $\Delta$F0]{\includegraphics[width=3.3in]{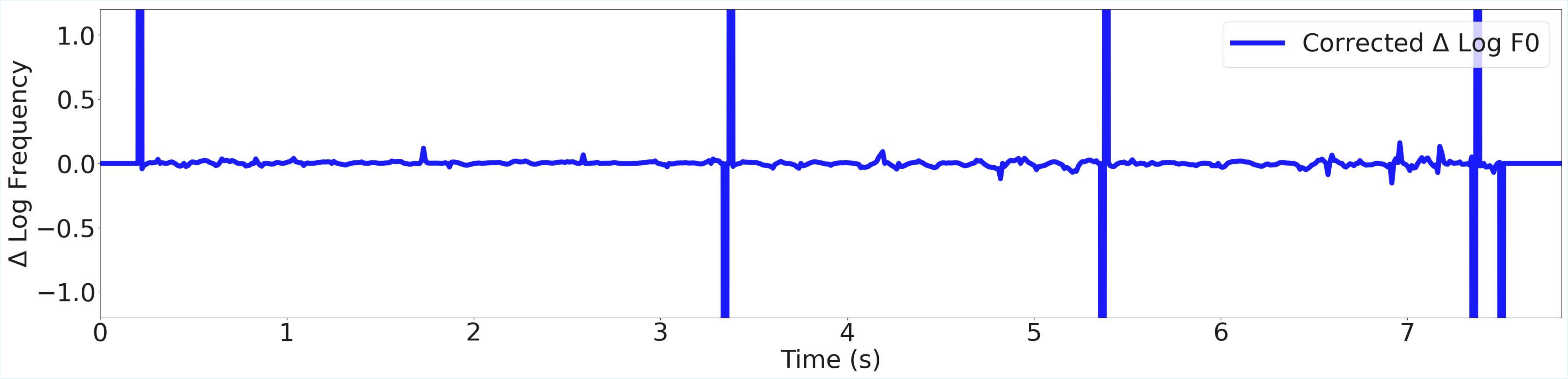}
\label{fig:shc_corrected}}

\caption{Demonstration of the SHC results. (a) visualizes the mel-spectrogram alongside the original F0 contour (red) and the corrected F0 contour (blue). (b) and (c) display the original $\Delta F0$ contour and the corrected $\Delta F0$ contour, respectively.}
\vspace{-3pt}
\end{figure}

\subsection{Timbre Style Conversion}
We expand the framework to enable the conversion of timbre styles in conjunction with pitch styles by incorporating a style lookup table (LUT). The independent disentanglement between these two attributes is structurally guaranteed by the modularity described in Section III.A. Because frame-level pitch variations are fully dictated by the explicitly provided F0 contour to the base SVC model, the global timbre LUT is forced to act as an information bottleneck, capturing only the average spectral envelope and phonation characteristics. This architectural constraint prevents high-frequency pitch-style information from leaking into the timbre representation, enabling the model to successfully combine diverse techniques (e.g., applying a vibrato pitch style over a belting timbre) without interference \cite{sintechsvs}. This allows the model to modify the timbre style and speaker identity by selecting respective target IDs. 
This approach successfully allows for independent control over pitch and timbre. However, we observed a problem where converting from a vocal fry audio results in an unnatural-sounding output due to its subharmonic characteristic. Such pitch jumps, caused by subharmonics, are also reflected in the F0 contour during conversion. To address this, we propose a Subharmonic Correction (SHC) algorithm based on the first-order difference of the F0 contour.

Our core assumption is that a large pitch jump or drop within a single frame (e.g. 10ms) is an artifact from the F0 estimation algorithm or a subharmonic component.
The algorithm begins by transforming the F0 contour into the log scale and then generating its first-order difference contour. Since an F0 contour presents pitch drop to half or a pitch jump to double respectively, as shown in Fig. \ref{fig:shc_original}, the absolute value of the difference in the log-F0 contour at that frame will be approximately $1$. We can detect and filter these subharmonic frames by applying a threshold to the difference contour, as shown in Fig. \ref{fig:shc_corrected}.
To reconstruct the corrected contour, we autoregressively accumulate the filtered first-order difference contour, starting from the initial log-F0 value. To anchor the reconstructed contour and minimize overall error, we scale the entire corrected contour based on the mode value between the original and corrected contour so that the entire corrected contour aligns well with the original F0 contour. If the mean value of the original contour is utilized to scale the contour, the corrected contour will be scaled at a lower range due to subharmonic in the original contour. Furthermore, it is important because we cannot always determine whether the source contour is in a subharmonic range or the original F0 range across its entire duration. Based on this process, as shown in Fig. \ref{fig:shc_mel}, our proposed SHC algorithm successfully removes pitch jumps and drops caused by subharmonic characteristics which occurred at the F0 contour.

\begin{algorithm}[t]
\caption{Subharmonic Correction (SHC)}\label{alg:alg1}
\begin{algorithmic}[h!]

\State \textbf{Input:} $f0$, $voiced$, $threshold$
\State \textbf{Output:} $\widetilde{f0}$

\State
\State $lf0 \gets \log_2(f0)$
\State $\Delta lf0 \gets lf0[1:]-lf0[:-1]$
\State
\For{$i=1,2,\ldots,T-1$ }
\If {$|\Delta lf0_i|  > threshold$ \\
$\quad\quad \textbf{and}$ $voiced_i=false$ \\
$\quad\quad \textbf{and}$ $voiced_{i+1}=false$}
        \State $\Delta lf0_i \gets 0$ 
\EndIf
\EndFor

\State
\State $\widetilde{lf0} \gets \text{new\_array\_of\_size(T)}$
\State $\widetilde {lf0}_1 \gets lf0_1$
\For{$i=2,3,\ldots,T$ }
\If {$voiced_i=false $\\
$\quad\quad \textbf{and}$ $voiced_{i+1}=false$}
\State $\widetilde {lf0}_{i+1} \gets \widetilde {lf0}_i + lf0_{i-1}$  
\Else
    \State $\widetilde {lf0}_{i+1} \gets \widetilde {lf0}_i$
\EndIf
\EndFor

\State

\State $scale \gets \textbf{mode}(lf0-\widetilde {lf0})$
\State $\widetilde {lf0} \gets \widetilde {lf0}+ scale $
\State $\widetilde {f0} \gets \textbf{pow}(\widetilde {lf0},2)$
\State \Return $\widetilde {f0}$

\end{algorithmic}
\label{algo1}
\end{algorithm}

\section{Experiments}
\subsection{Dataset}
As a general preprocessing step, all audio samples were downsampled to 24 kHz and segmented into 10 seconds to ensure training efficiency. 

The VocalSet\cite{vocalset} consists of 20 singers, including 11 males and 9 females. It contains 17 different singing styles sung in contexts of scale, arpeggios, and excerpts. For the pitch style conversion task, we selected samples with straight and vibrato singing styles. The test set consists of 80 samples, with 4 samples per style for each speaker, and all remaining samples were used for the training set. The same test set is used to evaluate the zero-shot pitch style conversion task.
For the timbre style conversion task, we selected 4 styles: straight, belt, breathy, and vocal fry. The test set comprises 160 samples, containing 2 samples per style for each speaker, with the rest of the data allocated to the training set.

The GTSinger dataset\cite{gtsinger} contains singing voice recordings from 20 singers across 9 languages and 6 singing styles. 
For the zero-shot pitch style conversion task, we utilized the control and vibrato styles from the vibrato group of 
English to train the pitch style encoder. We relabeled the control style as straight. We utilize audio samples across all english speakers to train the pitch style converter. We randomly selected $80 \%$ of the dataset for the training set.

\subsection{Feature Extraction}
We set the hop size to 256, the FFT size to 1024, and the window size to 1024 for mel-spectrogram extraction with 100 mel-frequency bins. We extracted an F0 contour using RMVPE \cite{rmvpe} and an energy contour from the root mean square (RMS) value of each frame, using the same hop size for both to ensure temporal alignment with the mel-spectrogram.
To represent linguistic content, we downsampled the audio to 16 kHz and processed it with a pretrained HuBERT-soft encoder \cite{softvc} to obtain semantic units. Then, we upsampled these units via linear interpolation to match the frame rate of the mel-spectrogram. 
During inference, we shifted the F0 contour to the mean F0 of the target speaker, and we decomposed the F0 contour into low- and high-frequency components using a DWT with the Daubechies-10 mother wavelet. We set the cutoff level of the DWT to 4.

\subsection{Implementation Details}
We trained the SVC model for 200k steps with a batch size of 128. The pitch style converter and energy style converter were trained with a batch size of 256 for 100k steps and 60k steps respectively. The dimension of timbre style embedding is set to 256. The remaining settings of the SVC, style converters, and discriminators are the same as VibE-SVC \cite{vibesvc}.
For waveform generation, we utilized a pretrained BigVGAN\footnote{\url{https://huggingface.co/nvidia/bigvgan_v2_24khz_100band_256x}} \cite{bigvgan} vocoder.
For training ZSC, the pitch style converter and the energy style converter are trained for 100k steps with a batch size of 256. The architecture of the high-frequency style encoder follows the mel style encoder of Meta-style speech \cite{metastylespeech}, except that its input channel is set to 1.

\subsection{Baseline Models}
We adopted the same set of baseline models used in the VibE-SVC. These baselines, all of which are based on the DiffSVC \cite{diffsvc} architecture from the open-source project SoVITS-SVC\footnote{\url{https://github.com/svc-develop-team/so-vits-svc}}, consist of three variants: an SVC model with an added style embedding, a second model that also incorporates a DWT to condition low-frequency F0 contour to the model, and a third model integrated with the performance style transfer (PST) \cite{pst} module with the official codes\footnote{\url{https://github.com/poohhsu/Singing-Performance-Style-Transfer}}. In addition, we adopt VibE-SVC as a baseline model. 

For the zero-shot style conversion task, we compared our model to two zero-shot SVC models and three SSC models.
The samples are generated from each official checkpoints:  NeuCoSVC2\footnote{\url{https://github.com/thuhcsi/NeuCoSVC/tree/NeuCoSVC2}}, Seed-SVC\footnote{\url{https://github.com/Plachtaa/seed-vc}}, Serenade\footnote{\url{https://github.com/lesterphillip/serenade}}, Vevo1.5\footnote{\url{https://github.com/open-mmlab/Amphion/tree/main/models/svc/vevosing}}, Vevo2\footnote{\url{https://github.com/open-mmlab/Amphion/tree/main/models/svc/vevo2}}. As Vevo1.5 and Vevo2 require textual transcripts for both source and reference audio, we provided the corresponding vowel character as the text input for vowel-only samples. For other non-vowel samples, we generated transcripts using a pretrained Whisper\footnote{\url{https://huggingface.co/openai/whisper-large-v3}}\cite{whisper} model. 

For the timbre style conversion task, we established two baseline models. The first is a modified VibE-SVC which incorporated a timbre style embedding without using the SHC algorithm. The second is the model which is conditioned on a chromagram to the diffusion decoder instead of the F0 contour. We hypothesize that using an octave-invariant chromagram will provide robustness against the subharmonic characteristic in a vocal fry style.

\subsection{Evaluation Metrics}
\subsubsection{Objective Evaluations}
We evaluate speaker similarity via Speaker Embedding Cosine Similarity (SECS) using an official checkpoint of WavLM\footnote{\url{https://huggingface.co/microsoft/wavlm-base-sv}}\cite{wavlm}. Intelligibility is evaluated via Word Error Rate (WER) using an official checkpoint of Qwen3-ASR\footnote{\url{https://huggingface.co/Qwen/Qwen3-ASR-1.7B}} \cite{qwen3-asr}. Style conversion performance is assessed through style accuracy and F0 Pearson Correlation (FPC). FPC evaluates zero-shot glissando F0 similarity. For style accuracy, we fine-tuned MERT\footnote{\url{https://huggingface.co/m-a-p/MERT-v1-330M}}\cite{mert} classifier with an official checkpoint. The pitch classifier,which was fine-tuned on straight and vibrato, utilizes Parselmouth \cite{parselmouth} pitch shift augmentation. The timbre classifier employs the pitch shift alongside WORLD-based \cite{world} vibrato extent scaling, which is adapted from VibE-SVC to ensure robustness against pitch fluctuations.

\subsubsection{Subjective Evaluations}
We conducted a 5-point naturalness mean opinion score (nMOS) and a 4-point similarity mean opinion score (sMOS) test to evaluate the naturalness and speaker similarity of the generated samples. The tests were conducted for 50 samples per model via Amazon MTurk. For each task, 20 people participated in the tests. The nMOS and sMOS are reported with $95\%$ confidence intervals.

\begin{table*}[t]
\caption{The results of the pitch style conversion task.\label{tab:pitch_tech_conversion}}
\renewcommand{\arraystretch}{1.6}
\centering
\resizebox{1.00\linewidth}{!}{
\begin{tabular}{l||cccc||cccc|cccc}
\hline
\multirow{2}{*}{\textbf{Model}} &
\multicolumn{4}{c||}{\textbf{Average}}&
\multicolumn{4}{c|}{\textbf{Vibrato} $\xrightarrow{}$ \textbf{Straight}} & 
\multicolumn{4}{c}{\textbf{Straight} $\xrightarrow{}$ \textbf{Vibrato}} \\

\cline{2-13}

& 
\multicolumn{1}{c}{\textbf{nMOS}} & 
\multicolumn{1}{c}{\textbf{sMOS}} &
\multicolumn{1}{c}{\textbf{SECS}} & 
\multicolumn{1}{c||}{\textbf{Acc}} &
\multicolumn{1}{c}{\textbf{nMOS}} & 
\multicolumn{1}{c}{\textbf{sMOS}} &
\multicolumn{1}{c}{\textbf{SECS}} & 
\multicolumn{1}{c|}{\textbf{Acc}} &
\multicolumn{1}{c}{\textbf{nMOS}} & 
\multicolumn{1}{c}{\textbf{sMOS}} &
\multicolumn{1}{c}{\textbf{SECS}} & 
\multicolumn{1}{c}{\textbf{Acc}} \\

\hline

\textbf{GT}
& 4.501 $\pm$ 0.046 & 3.420 $\pm$ 0.040 
& 0.829             & 0.988
& 4.485 $\pm$ 0.064 & 3.430 $\pm$ 0.059 
& 0.828             & 0.975
& 4.518 $\pm$ 0.065 & 3.410 $\pm$ 0.056 
& 0.830             & 1.000\\

\textbf{Vocoded} 
& 4.447 $\pm$ 0.045 & 3.446 $\pm$ 0.041 
& 0.824             & 0.988 
& 4.430 $\pm$ 0.063 & 4.445 $\pm$ 0.059 
& 0.823             & 0.975 
& 4.465 $\pm$ 0.065 & 3.447 $\pm$ 0.056 
& 0.826             & 1.000\\
\hline

\textbf{SoVITS + Style Emb }          
& \underline{3.944 $\pm$ 0.077} & \textbf{2.946 $\pm$ 0.068 }
& 0.779             & 0.160    
& \textbf{3.983 $\pm$ 0.105}    & \textbf{2.960 $\pm$ 0.100}
& 0.774             & 0.018 
& 3.903 $\pm$ 0.112 & 2.933 $\pm$ 0.093 
& \textbf{0.785}    & 0.301 \\

\textbf{SoVITS + Style Emb + Low F0}
& 3.924 $\pm$ 0.078 & 2.906 $\pm$ 0.068 
& 0.784             & 0.540    
& 3.809 $\pm$ 0.108 & 2.875 $\pm$ 0.098 
& 0.784             & 0.361  
& \textbf{4.044 $\pm$ 0.110}    & 2.933 $\pm$ 0.093 
& \textbf{0.785}    & 0.718  \\

\textbf{SoVITS + PST \cite{pst}} 
& 3.816 $\pm$ 0.090 & 2.909 $\pm$ 0.068 
& 0.771             & 0.488    
& 3.809 $\pm$ 0.129 & \underline{2.930 $\pm$ 0.099}
& 0.772             & \textbf{0.949} 
& 3.823 $\pm$ 0.124 & 2.890 $\pm$ 0.095 
& 0.771             & 0.028 \\

\textbf{VibE-SVC \cite{vibesvc}} 
& 3.924 $\pm$ 0.076 & \underline{2.937 $\pm$ 0.067 }
& \textbf{0.785}    & \underline{0.693}
& 3.957 $\pm$ 0.107 & 2.926 $\pm$ 0.097 
& \underline{0.788} & 0.636 
& 3.889 $\pm$ 0.108 & \underline{2.947 $\pm$ 0.092}
& 0.783             & \textbf{0.750} \\
\hline

\textbf{VibE-SVC2 (Ours)}
& \textbf{4.000 $\pm$ 0.077 }   & 2.932 $\pm$ 0.069 
& \textbf{0.785}                & \textbf{0.736} 
& \underline{3.966 $\pm$ 0.109 }& 2.879 $\pm$ 0.099 
& \textbf{0.789}                & \underline{0.725}
& \underline{4.035 $\pm$ 0.110 }& \textbf{2.980 $\pm$ 0.096}
& 0.782                         & \underline{0.747} \\
\hline
\end{tabular}}
\vspace{3pt}
\end{table*}
\begin{table*}[t]
\caption{The results of the zero-shot pitch style conversion task.}\label{tab:pitch_tech_conversion_zeroshot}
\renewcommand{\arraystretch}{1.6}
\centering

\resizebox{1.00\linewidth}{!}{
\begin{tabular}{l||cccc||cccc|cccc}
\hline
\multirow{2}{*}{\textbf{Model}} &
\multicolumn{4}{c||}{\textbf{Average}}&
\multicolumn{4}{c|}{\textbf{Vibrato $\xrightarrow{}$ Straight}}&
\multicolumn{4}{c}{\textbf{Straight $\xrightarrow{}$ Vibrato}} \\

\cline{2-13}

& 
\multicolumn{1}{c}{\textbf{nMOS}}&
\multicolumn{1}{c}{\textbf{sMOS}}&
\multicolumn{1}{c}{\textbf{SECS}}&
\multicolumn{1}{c||}{\textbf{Acc}}&
\multicolumn{1}{c}{\textbf{nMOS}}&
\multicolumn{1}{c}{\textbf{sMOS}}&
\multicolumn{1}{c}{\textbf{SECS}}&
\multicolumn{1}{c|}{\textbf{Acc}}&
\multicolumn{1}{c}{\textbf{nMOS}}&
\multicolumn{1}{c}{\textbf{sMOS}}&
\multicolumn{1}{c}{\textbf{SECS}}&
\multicolumn{1}{c}{\textbf{Acc}} \\
\hline

\textbf{Seed-SVC \cite{seedvc}}
& 3.916 $\pm$ 0.059     & \underline{3.297 $\pm$ 0.043}
& 0.722             & 0.215  

& 3.927 $\pm$ 0.081     & \textbf{3.318 $\pm$ 0.061}
& 0.726                 & 0.076

& \underline{3.904 $\pm$ 0.086}     & 3.277 $\pm$ 0.061
& \textbf{0.810}                 & 0.163 \\

\textbf{NeuCoSVC2 \cite{neuco-svc}}
& \textbf{3.971 $\pm$ 0.060}    & \textbf{3.314 $\pm$ 0.043}
& \textbf{0.803}              & 0.133

& \textbf{4.000 $\pm$ 0.084}     & \underline{3.304 $\pm$ 0.063}
& \textbf{0.795}                 & 0.102

& \textbf{3.942 $\pm$ 0.086}     & \textbf{3.324 $\pm$ 0.060}
& 0.718                 & 0.354 \\
\hline

\textbf{Serenade \cite{serenade}} 
& 3.848 $\pm$ 0.067     & 3.228 $\pm$ 0.050
& 0.771                 & 0.121

& 3.906 $\pm$ 0.094     & 3.223 $\pm$ 0.072
& 0.768                 & 0.184

& 3.791 $\pm$ 0.097     & 3.233 $\pm$ 0.068
& 0.773                 & 0.057 \\

\textbf{Vevo1.5 \cite{vevo2}} 
& 3.841 $\pm$ 0.063     & 3.250 $\pm$ 0.046
& 0.775                 & 0.342

& 3.855 $\pm$ 0.091     & 3.262 $\pm$ 0.066
& 0.778                 & 0.195

& 3.827 $\pm$ 0.087     & 3.238 $\pm$ 0.065
& 0.772                 & \underline{0.488} \\

\textbf{Vevo2 \cite{vevo2}} 
& \underline{3.917 $\pm$ 0.060}     & 3.274 $\pm$ 0.048
& \underline{0.790}                 & \underline{0.427}

& \underline{3.937 $\pm$ 0.083}     & 3.292 $\pm$ 0.070
& 0.778                 & \underline{0.366}

& 3.897 $\pm$ 0.086     & 3.255 $\pm$ 0.065
& \underline{0.802}                 & 0.390 \\
\hline

\textbf{VibE-SVC2-ZSC (Ours)}
& 3.876 $\pm$ 0.064     & 3.271 $\pm$ 0.046
& 0.787                 & \textbf{0.614}

& 3.874 $\pm$ 0.089     & 3.259 $\pm$ 0.062
& \underline{0.790}                 & \textbf{0.515}

& 3.877 $\pm$ 0.092     & \underline{3.283 $\pm$ 0.069}
& 0.785                 & \textbf{0.713} \\
\hline
\end{tabular}}
\end{table*}
\begin{figure*}[!t]
\centering
\subfloat[Source]{\includegraphics[width=1.1in, height=0.75in]{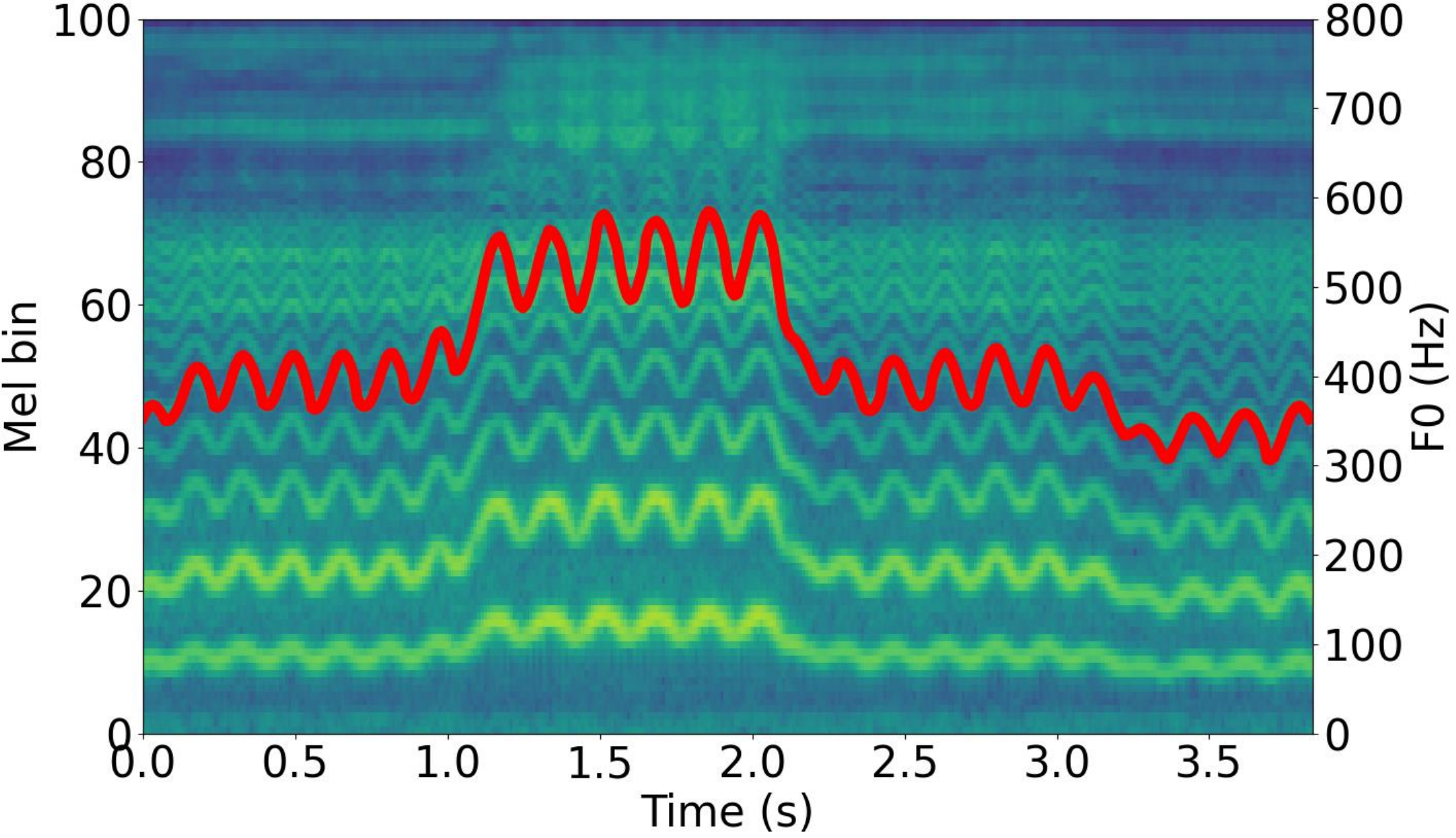}%
\label{fig:zs_vib2str_src}}
\hfil
\subfloat[Reference]{\includegraphics[width=1.1in, height=0.75in]{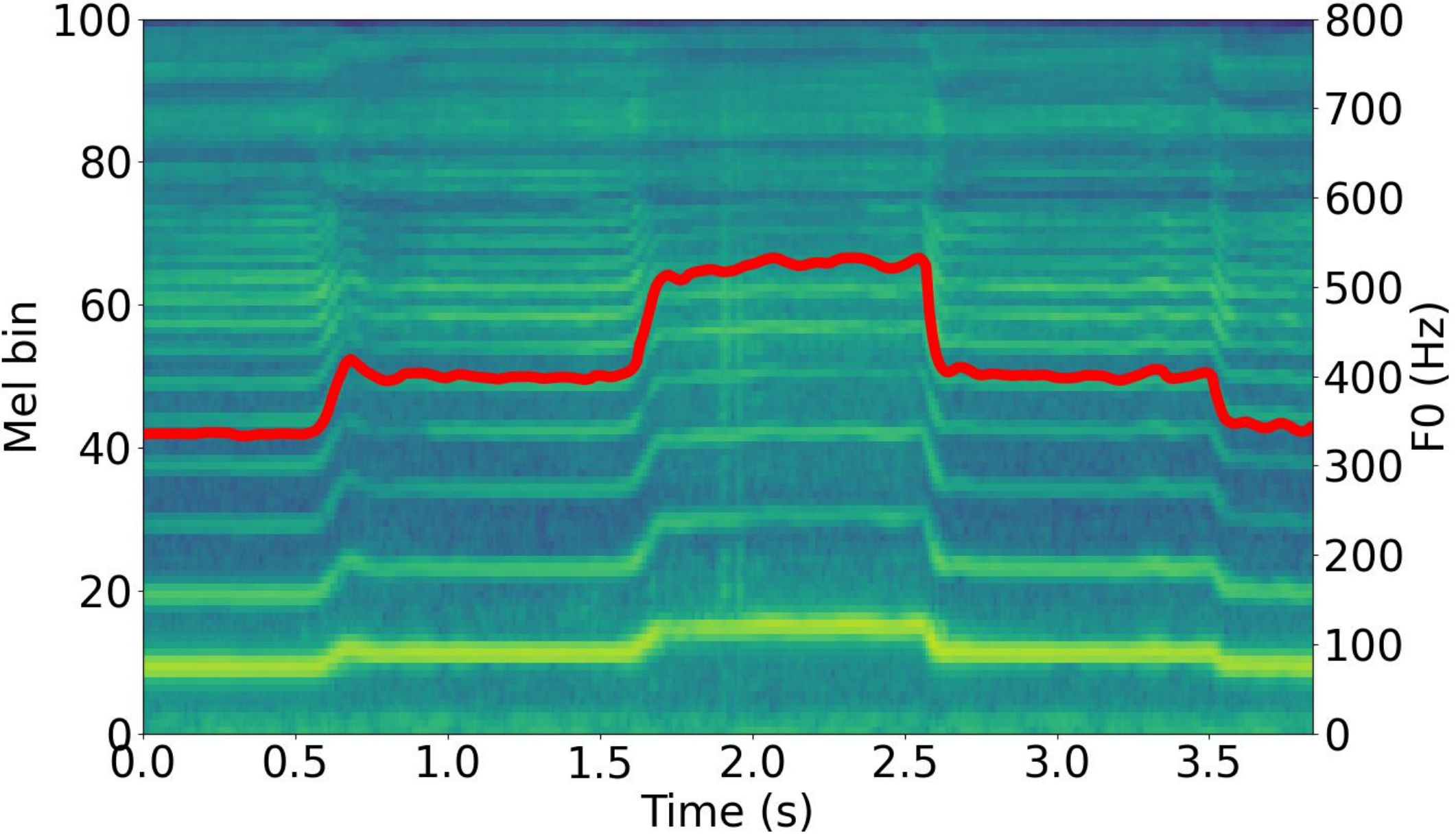}%
\label{fig:zs_vib2str_ref}}
\hfil
\subfloat[Serenade]{\includegraphics[width=1.1in, height=0.75in]{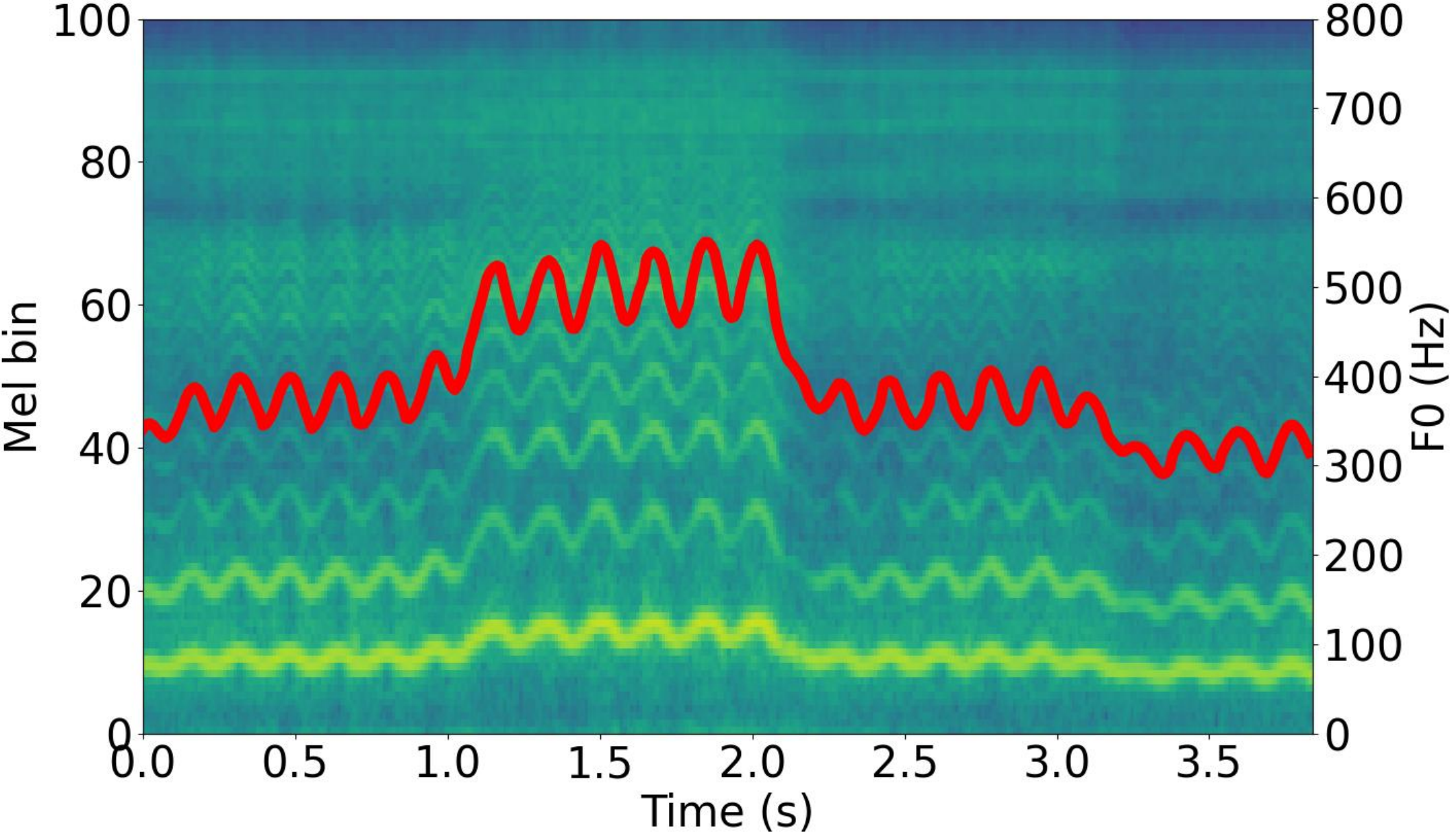}%
\label{fig:zs_vib2str_serenade}}
\hfil
\subfloat[Vevo1.5]{\includegraphics[width=1.1in, height=0.75in]{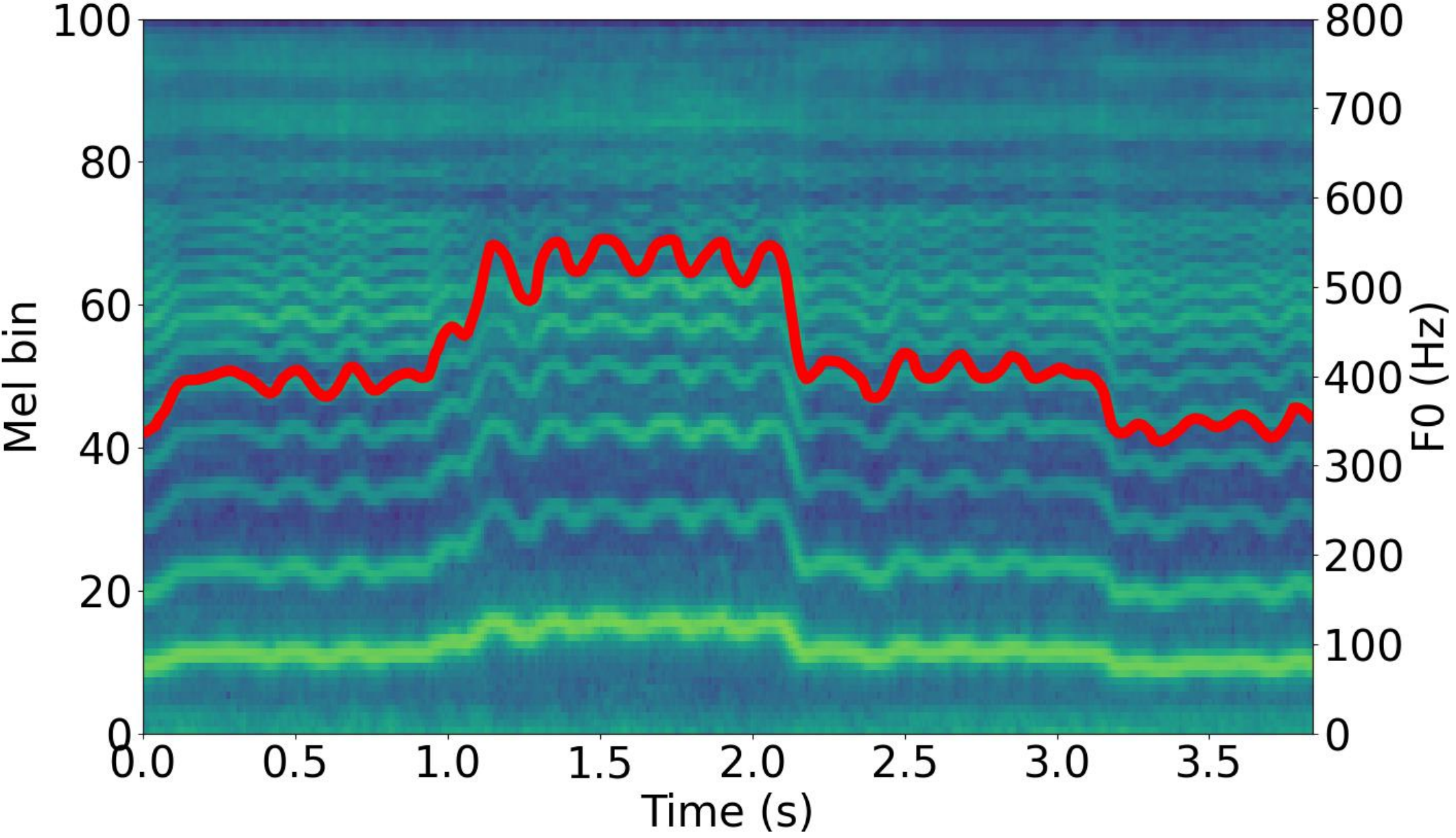}%
\label{fig:zs_vib2str_vevosing}}
\hfil
\subfloat[Vevo2]{\includegraphics[width=1.1in, height=0.75in]{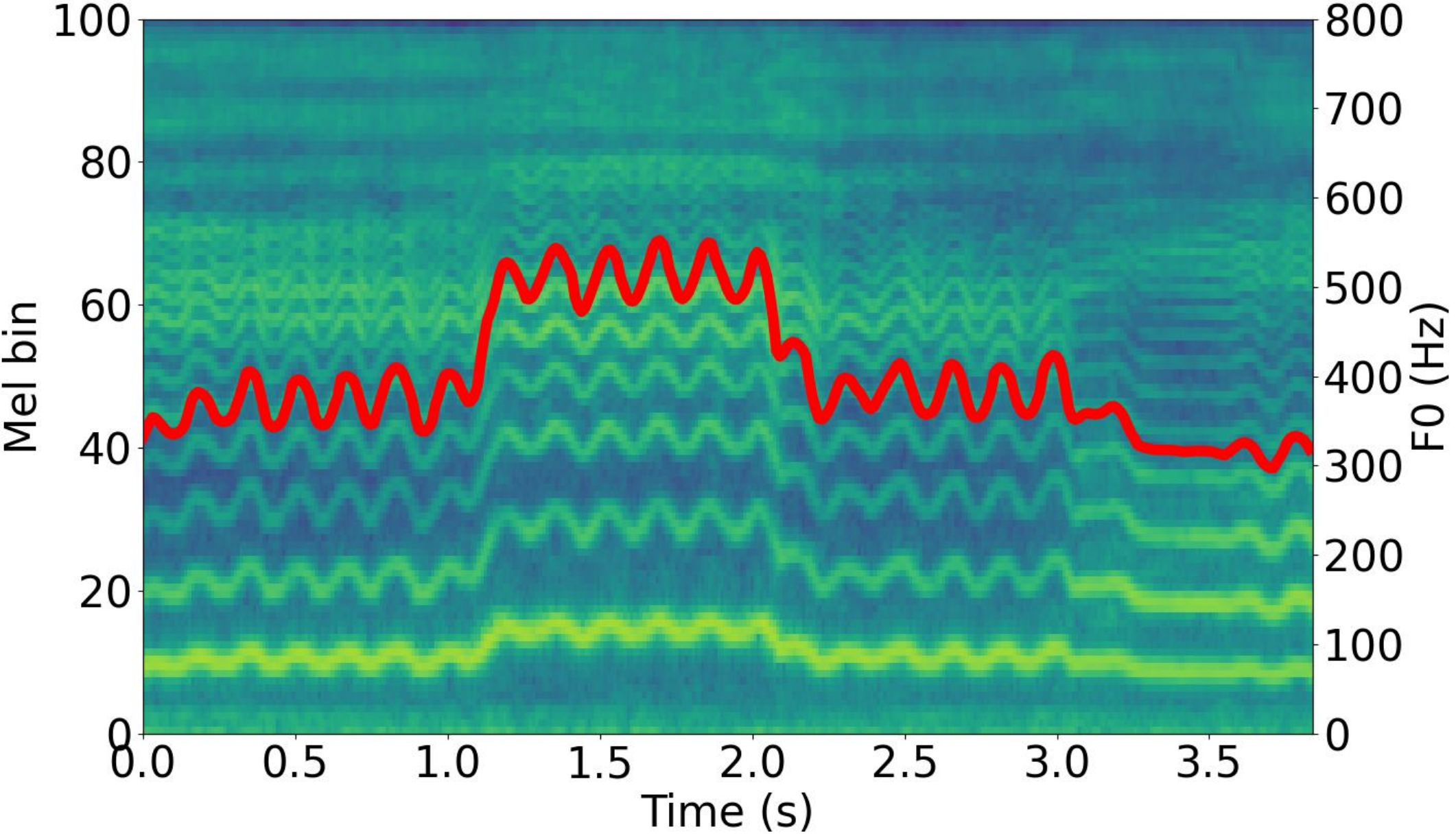}%
\label{fig:zs_vib2str_vevo2}}
\hfil
\subfloat[VibE-SVC2-ZSC]{\includegraphics[width=1.1in, height=0.75in]{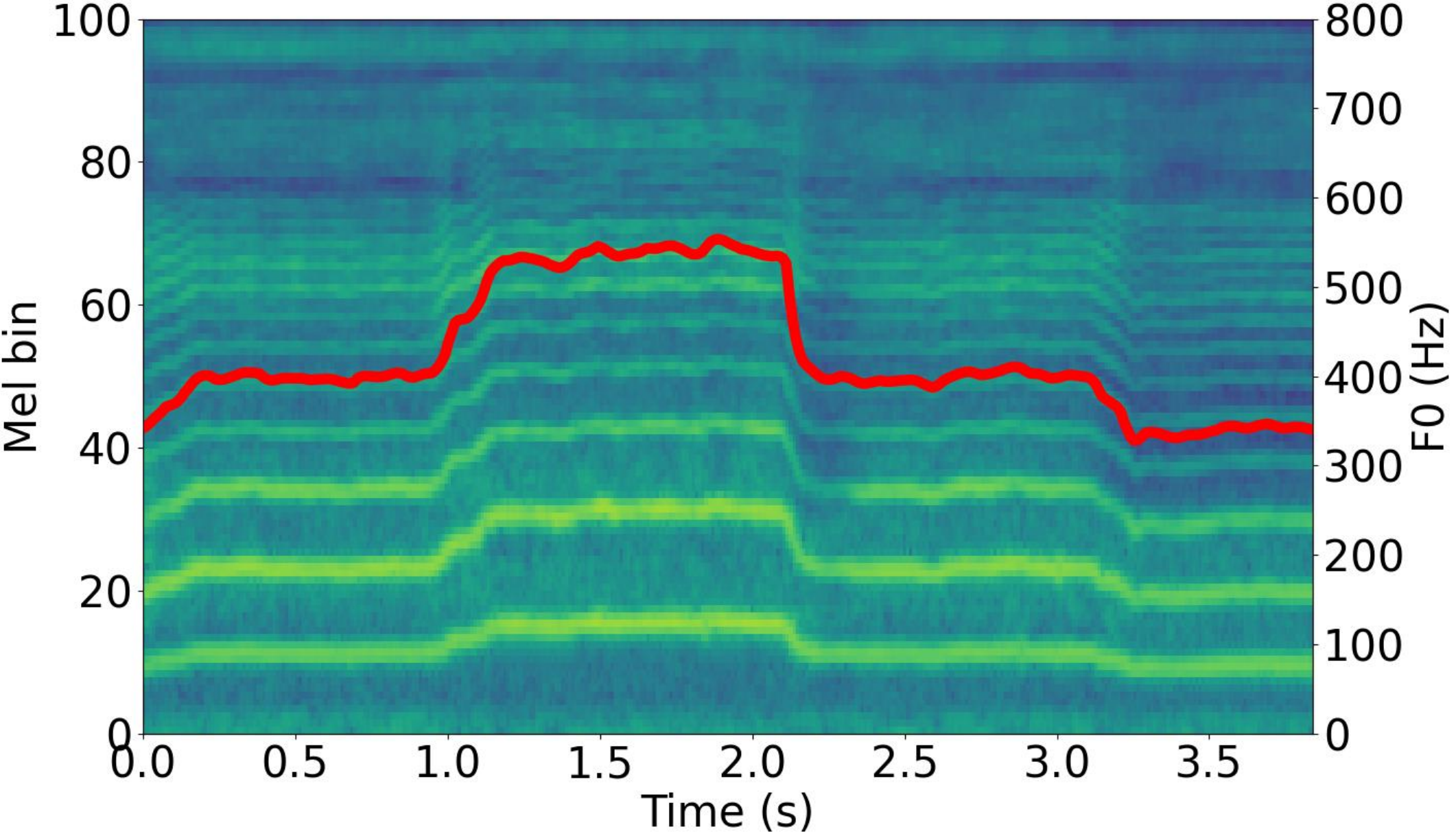}%
\label{fig:zs_vib2str_vibesvc2}}

\subfloat[Source]{\includegraphics[width=1.1in, height=0.75in]{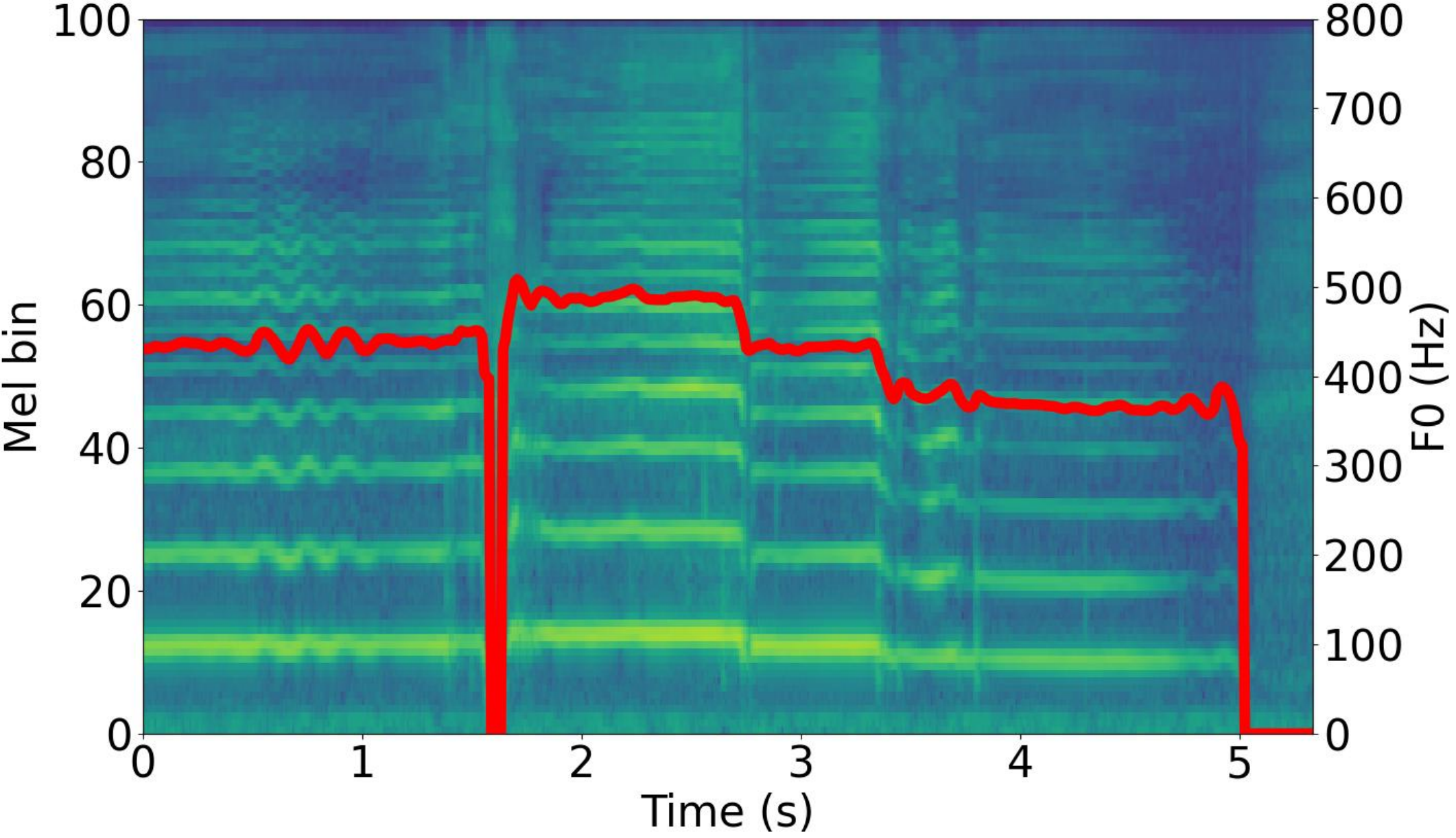}
\label{fig:zs_str2vib_src}}
\hfil
\subfloat[Reference]{\includegraphics[width=1.1in, height=0.75in]{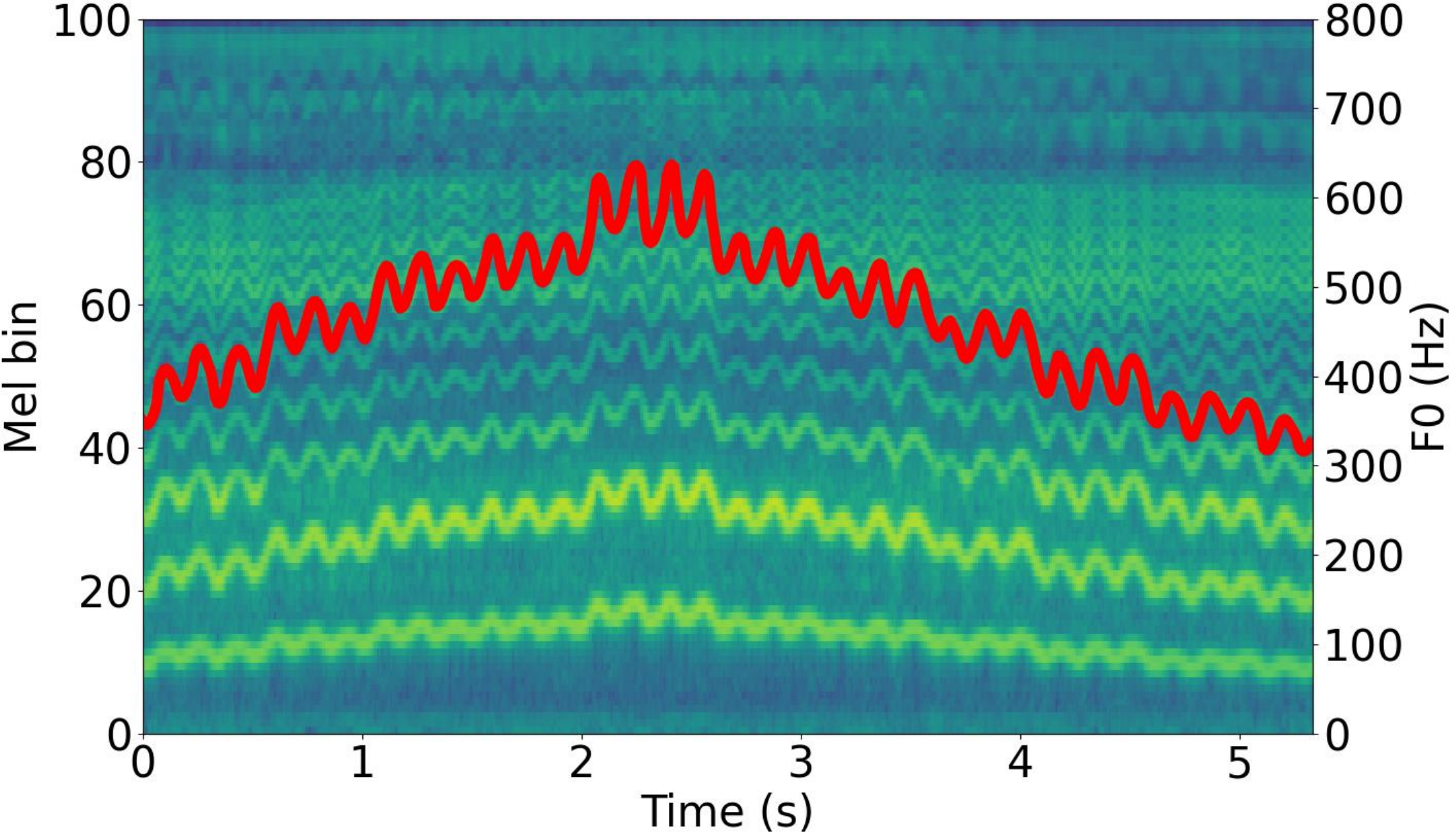}%
\label{fig:zs_str2vib_ref}}
\hfil
\subfloat[Serenade]{\includegraphics[width=1.1in, height=0.75in]{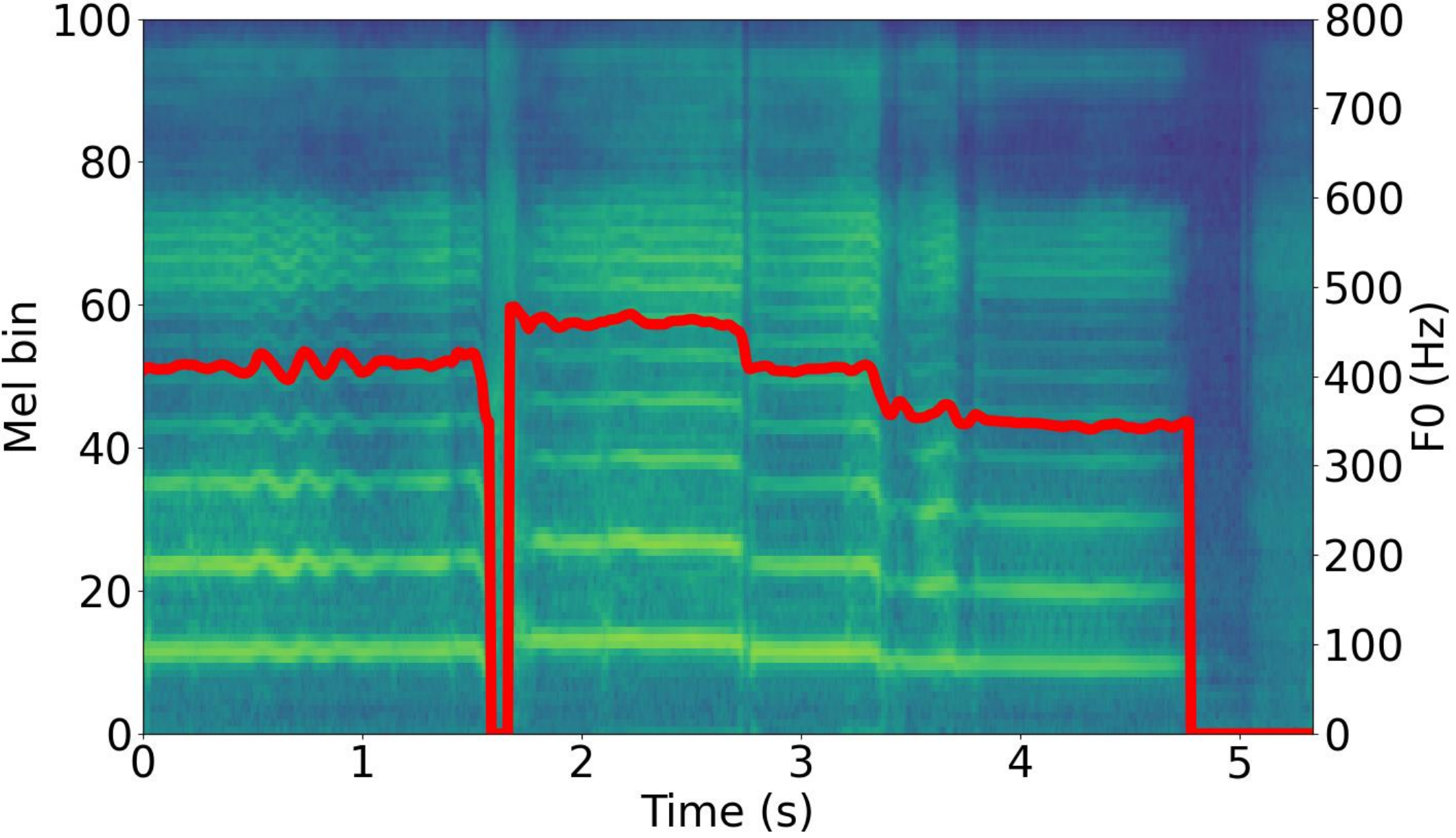}
\label{fig:zs_str2vib_serenade}}
\hfil
\subfloat[Vevo1.5]{\includegraphics[width=1.1in, height=0.75in]{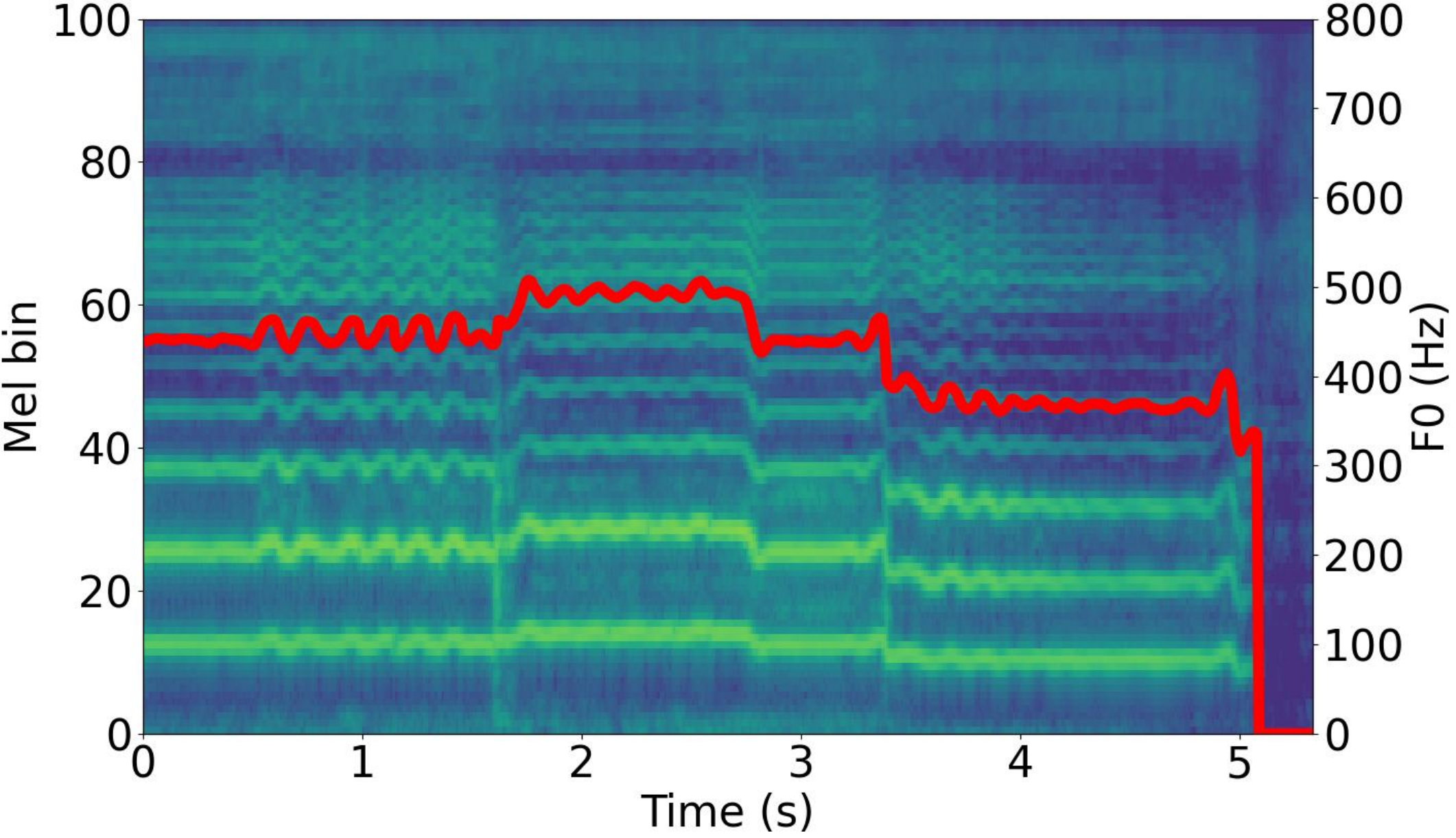}%
\label{fig:zs_str2vib_vevosing}}
\hfil
\subfloat[Vevo2]{\includegraphics[width=1.1in, height=0.75in]{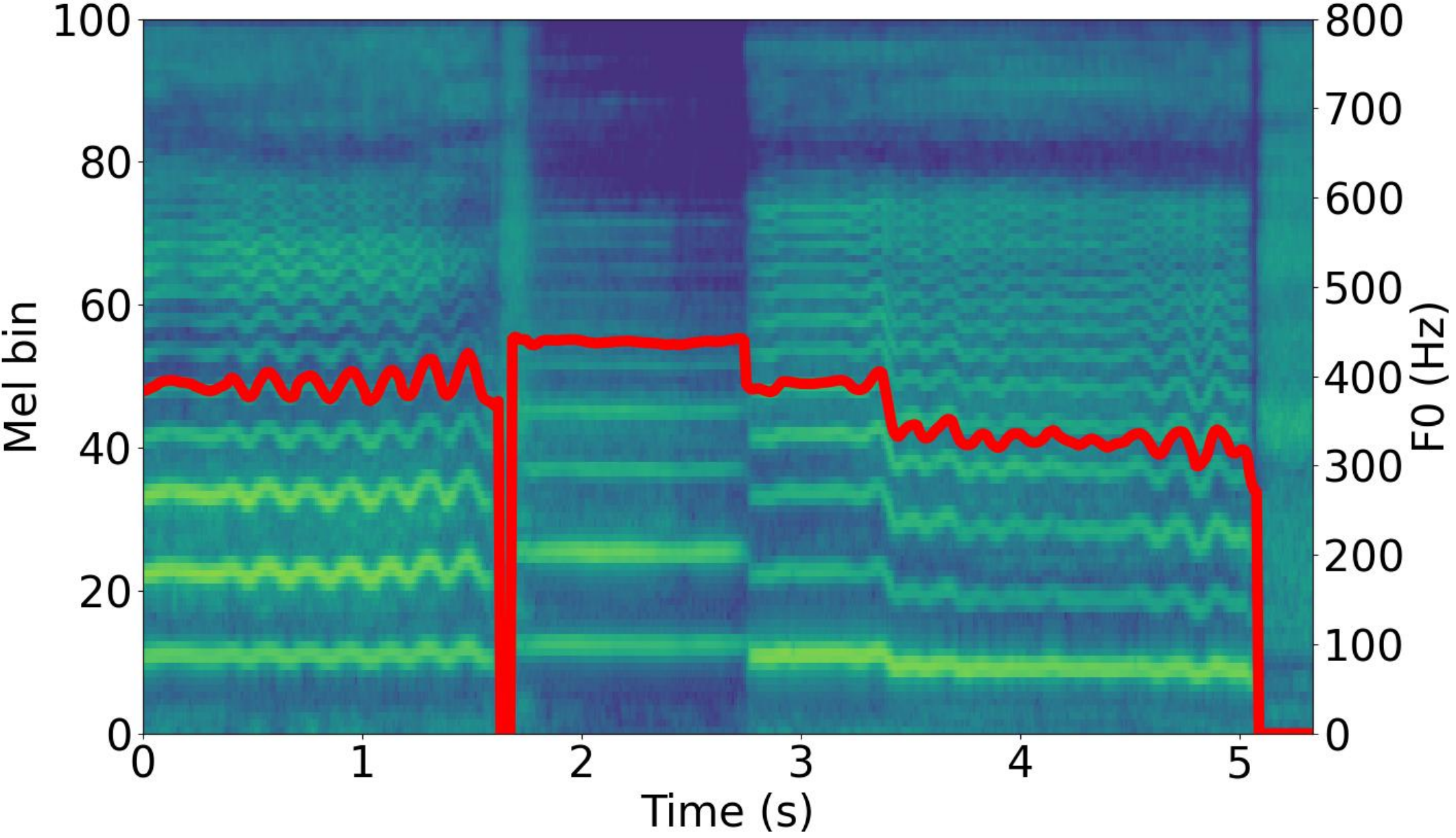}%
\label{fig:zs_str2vib_vevo2}}
\hfil
\subfloat[VibE-SVC2-ZSC]{\includegraphics[width=1.1in, height=0.75in]{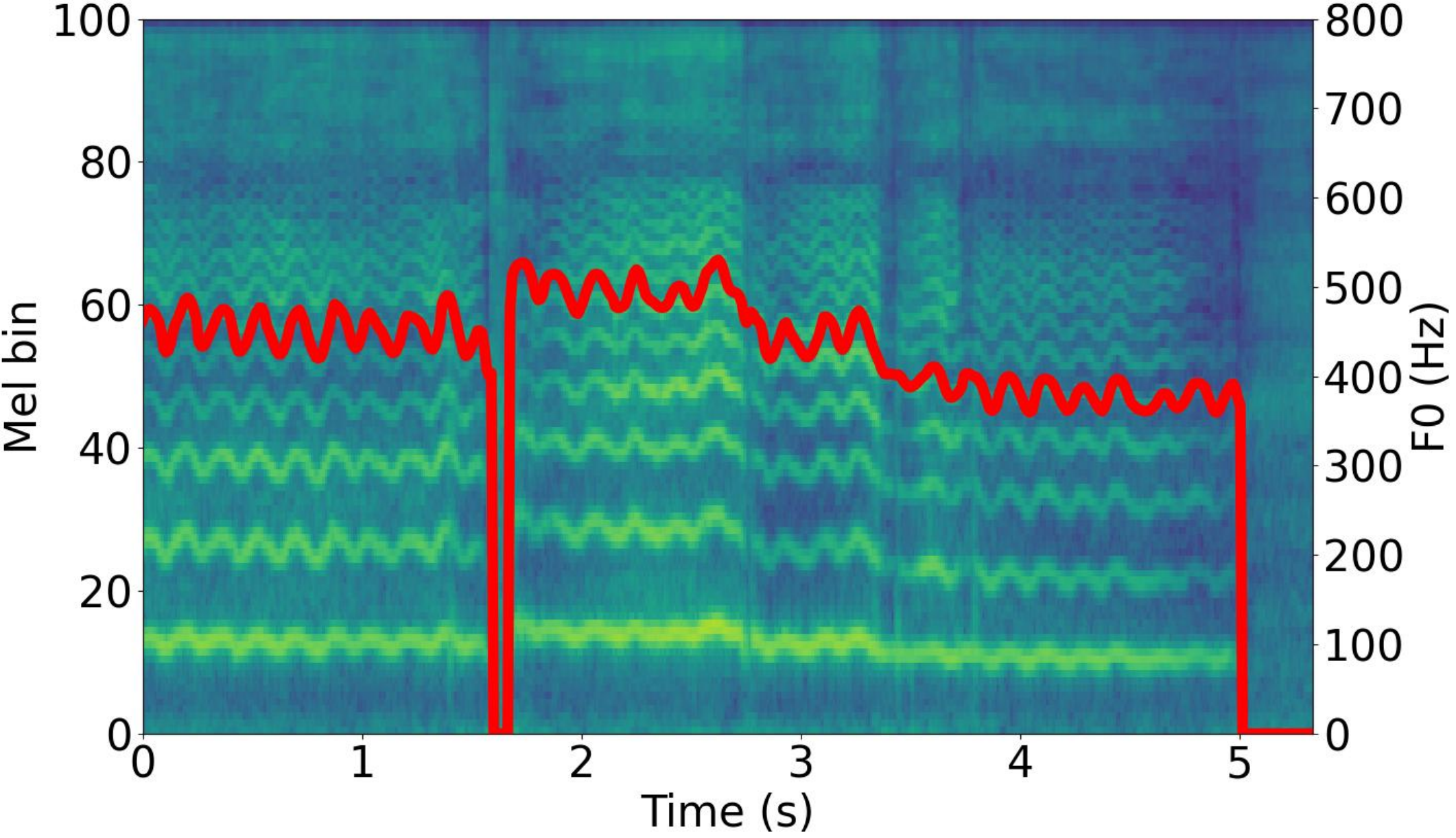}%
\label{fig:zs_str2vib_vibesvc2}}
\caption{Comparison of mel-spectrograms from the zero-shot pitch style conversion task. The results show the Vibrato $\xrightarrow{}$ Straight conversion (top) and Straight $\xrightarrow{}$ Vibrato conversion (bottom).}

\label{fig:zero_shot}
\end{figure*}

\section{Results}
\subsection{Performance of Pitch Style Conversion}
As shown in Table \ref{tab:pitch_tech_conversion}, our proposed energy style converter significantly improves the average conversion performance of the pitch style, especially in the Vibrato $\xrightarrow{}$ Straight conversion. The result shows that VibE-SVC2 effectively disentangles source style information, unlike in the baseline VibE-SVC where vibrato fluctuation is still in the energy contour. For the other baseline models, they exhibit biased performance, resulting in high performance for one direction style conversion while showing low performance at the other direction, especially in the SoVITS with PST model. This means that the model fails to disentangle the style information from the source audio. However, our proposed model achieves the best average score in style accuracy. This indicates that VibE-SVC2 disentangles pitch styles more effectively than the baseline models. VibE-SVC2 also achieves the best performance in naturalness and comparable performance in speaker similarity. VibE-SVC2 shows the best average speaker similarity in SECS and slightly lower performance within the range of confidence interval in the sMOS. In summary, VibE-SVC2 improves style conversion performance with comparable naturalness and speaker similarity performance.

\begin{figure*}[!t]
\centering
\subfloat[Source (Straight)]{\includegraphics[width=1.3in, height=0.8in]{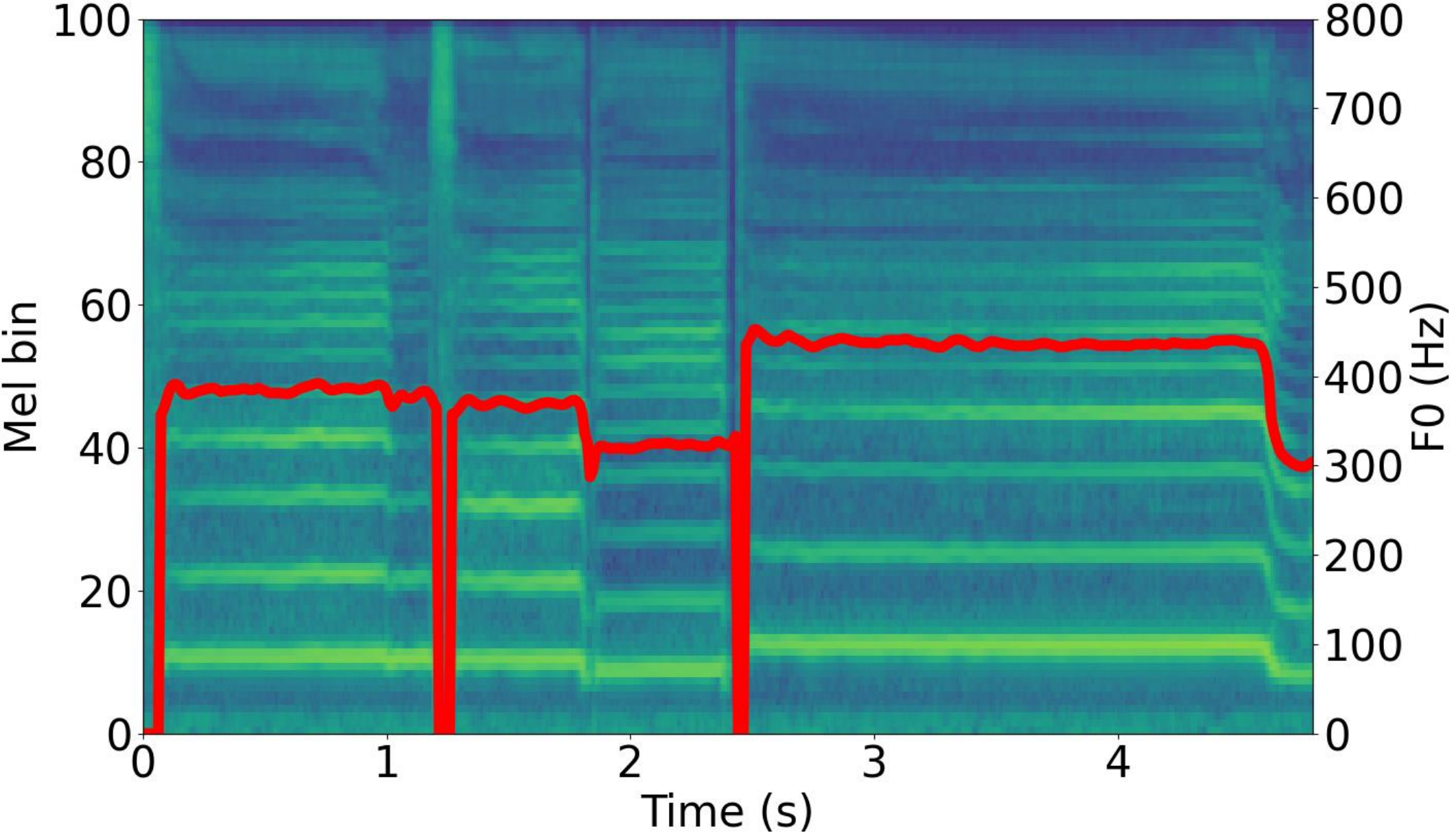}
\label{rate_control_source}}
\hfil
\subfloat[$\alpha=1.0$, $\beta=0.1$]{\includegraphics[width=1.3in, height=0.8in]{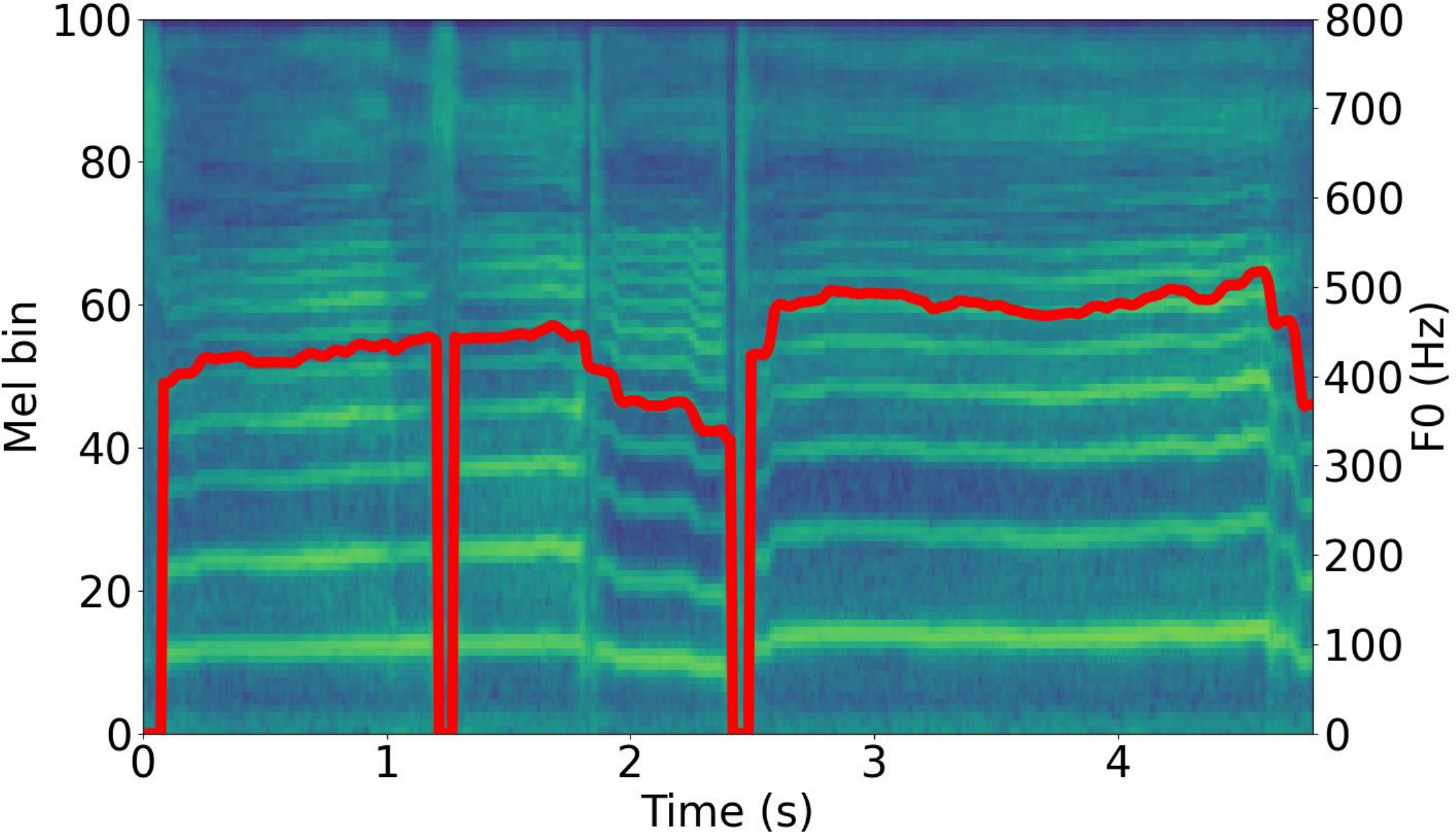}%
\label{fig:rate_control_ext1.0_rate_0.1}}
\hfil
\subfloat[$\alpha=1.0$, $\beta=0.5$]{\includegraphics[width=1.3in, height=0.8in]{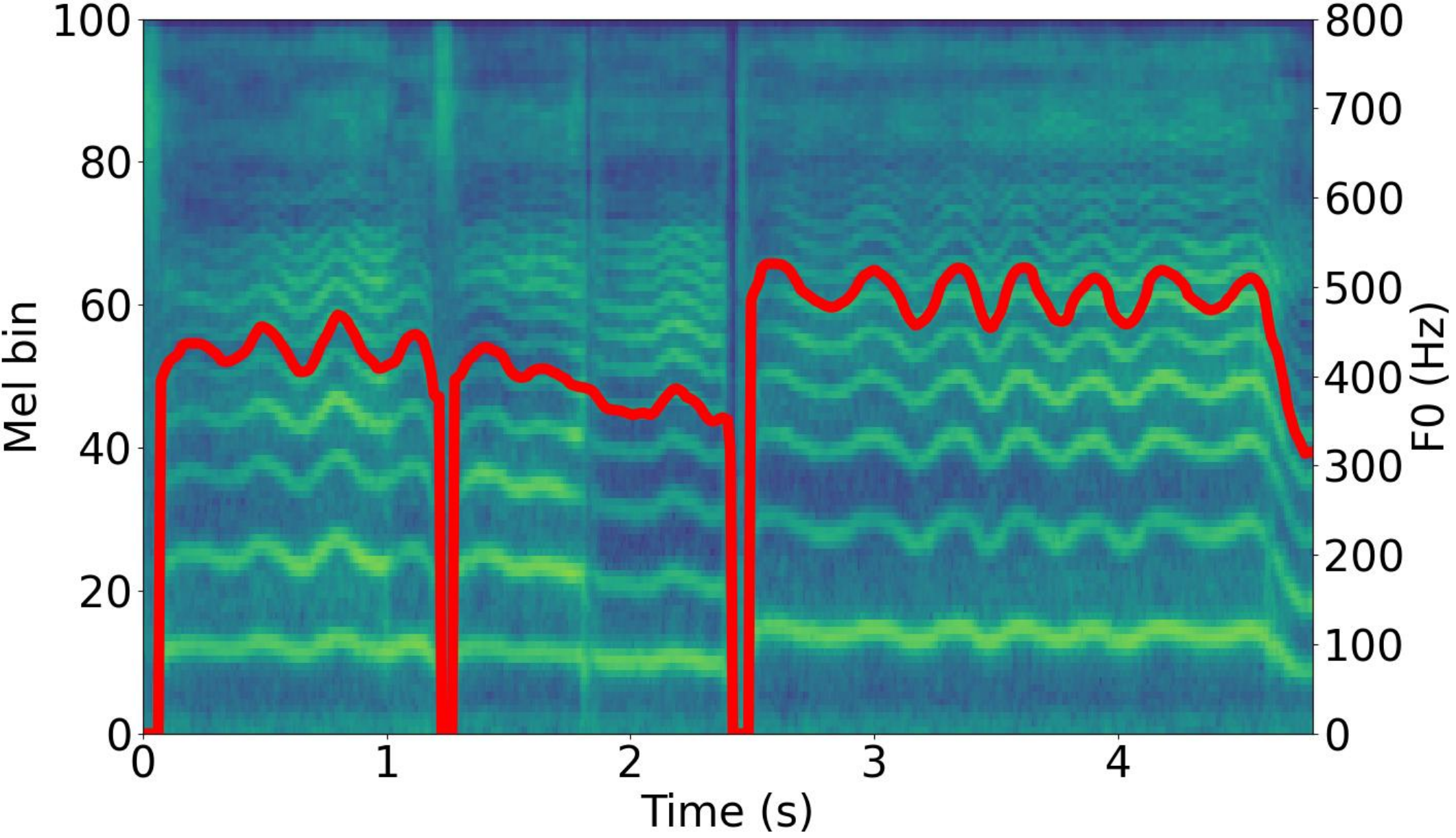}%
\label{fig:rate_control_ext1.0_rate_0.5}}
\hfil
\subfloat[$\alpha=1.0$, $\beta=1.5$]{\includegraphics[width=1.3in, height=0.8in]{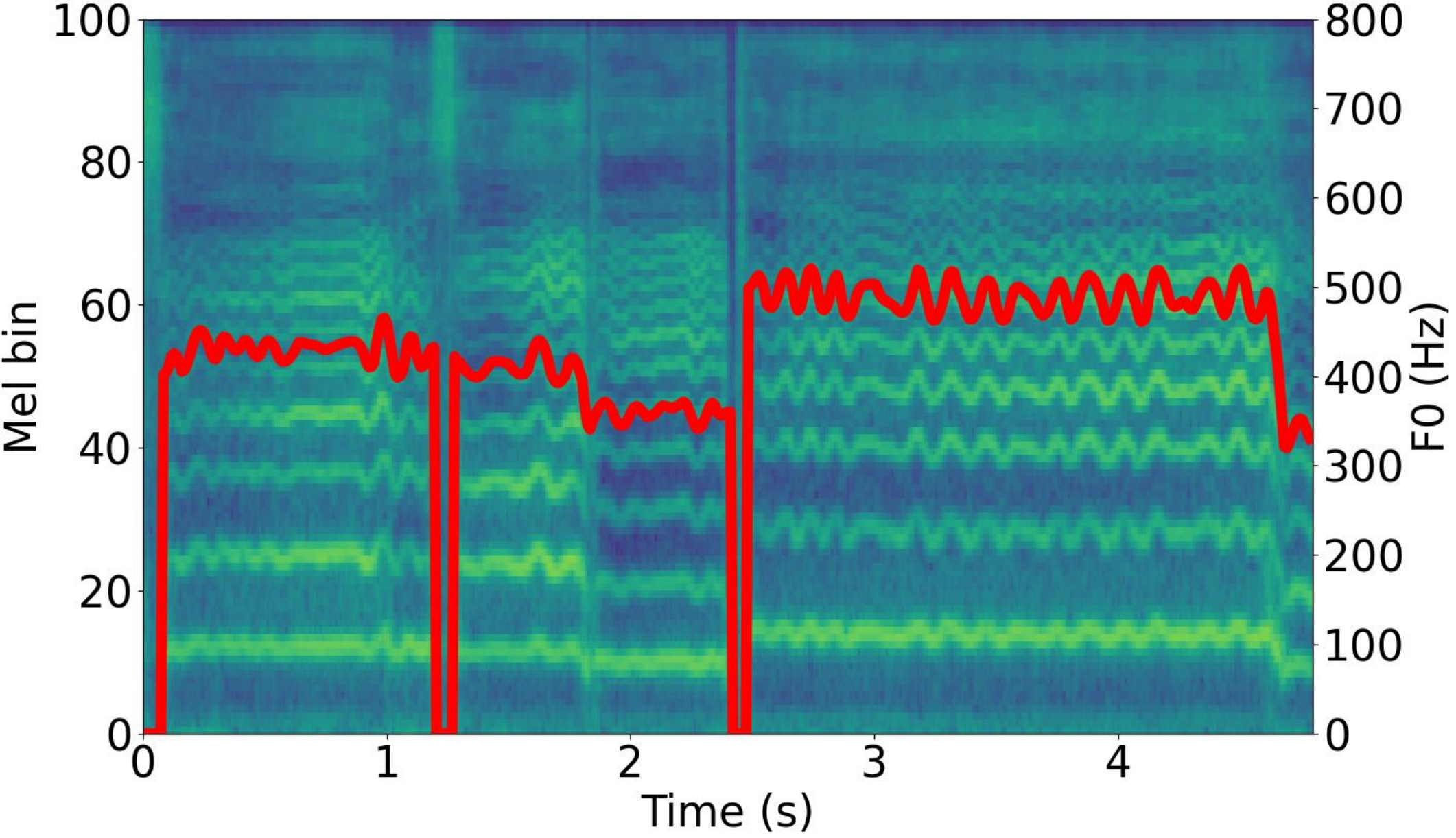}%
\label{fig:rate_control_ext1.0_rate_1.5}}
\hfil
\subfloat[$\alpha=1.0$, $\beta=2.0$]{\includegraphics[width=1.3in, height=0.8in]{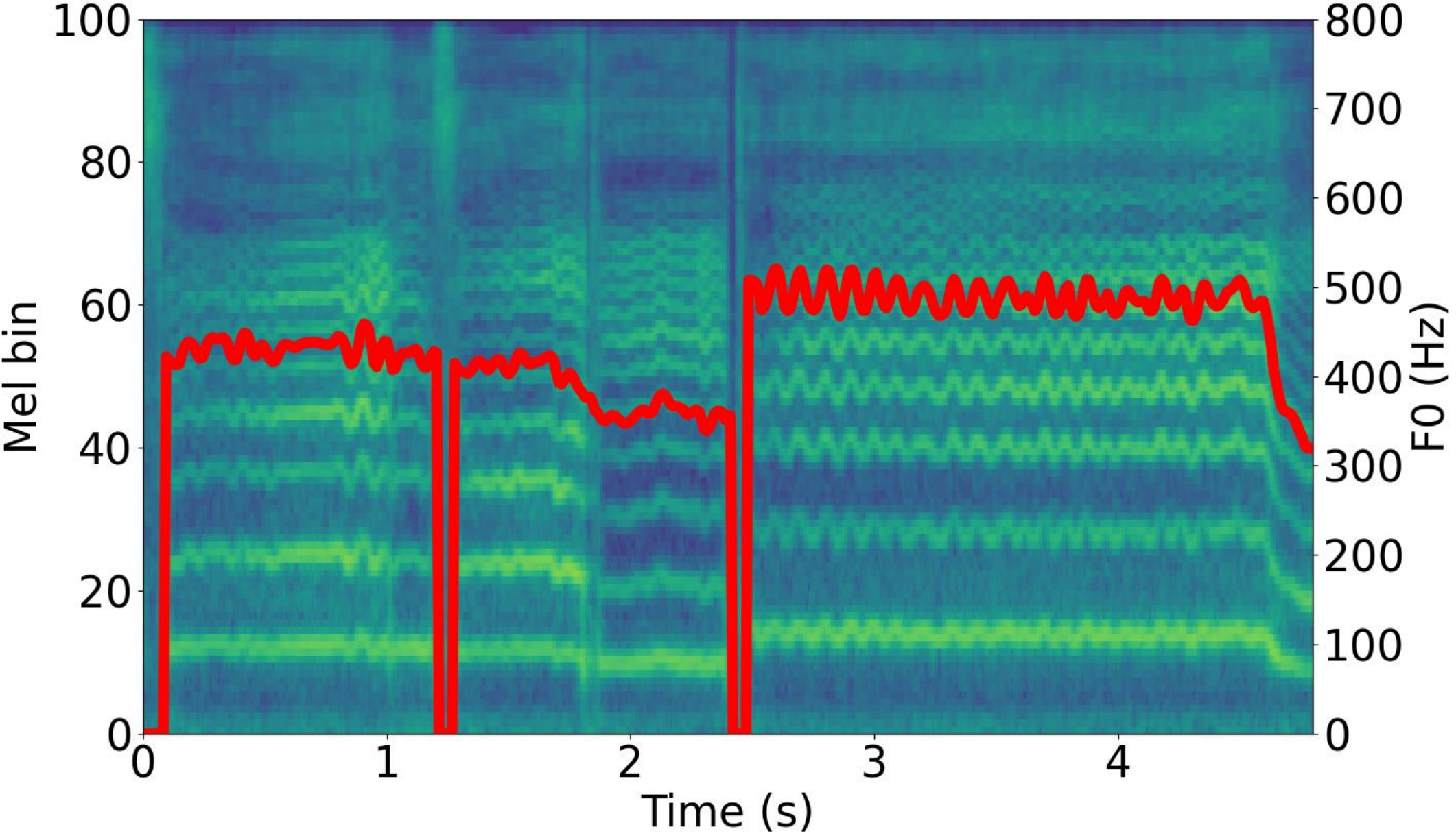}%
\label{fig:rate_control_ext1.0_rate_2.0}}

\subfloat[Straight $\xrightarrow{}$ Vibrato]{\includegraphics[width=1.3in, height=0.8in]{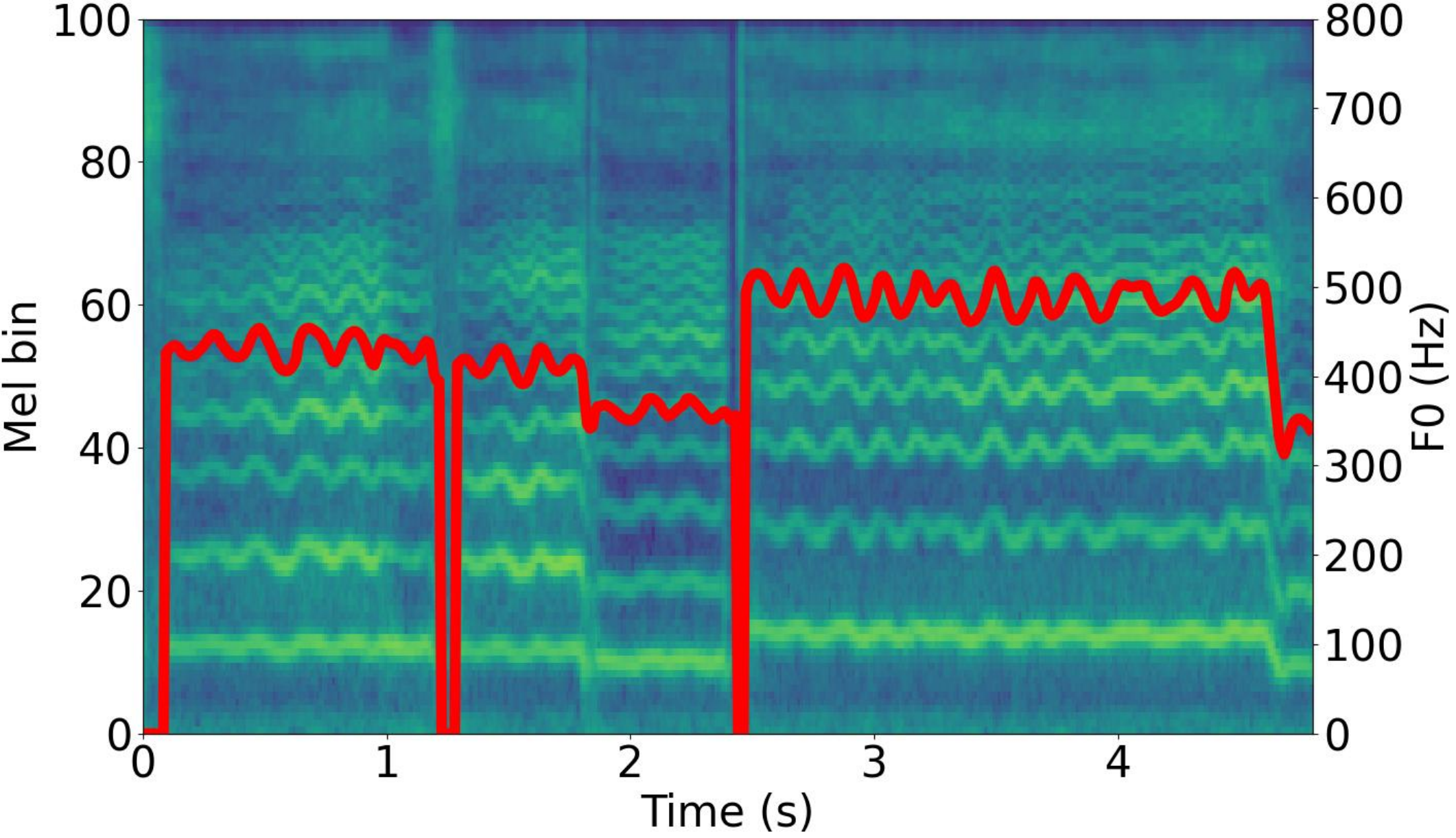}%
\label{fig:rate_control_ext1.0_rate_1.0}}
\hfil
\subfloat[$\alpha=0.5$, $\beta=0.5$]{\includegraphics[width=1.3in, height=0.8in]{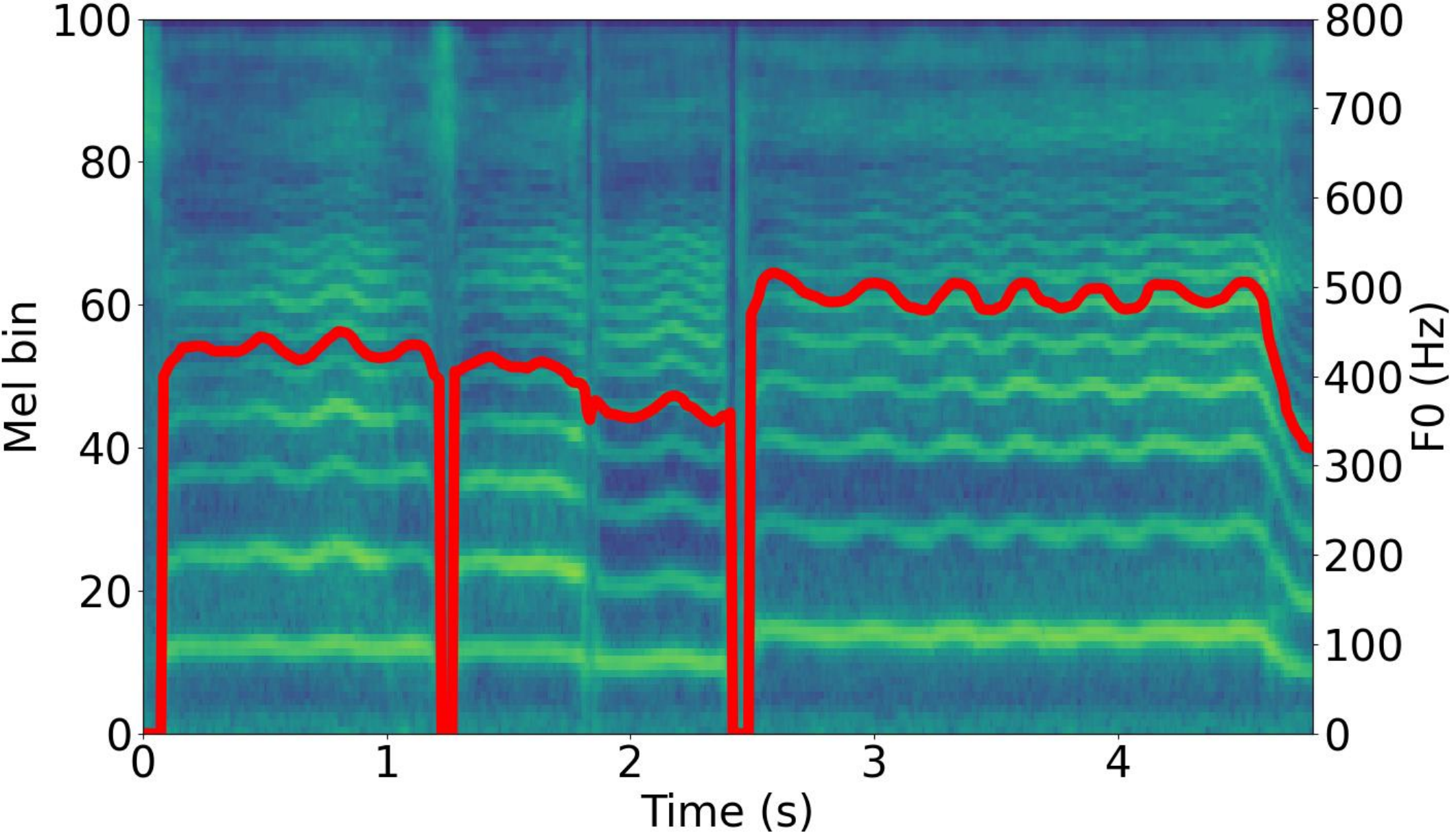}%
\label{fig:rate_control_ext0.5_rate_0.5}}
\hfil
\subfloat[$\alpha=1.5$, $\beta=0.5$]{\includegraphics[width=1.3in, height=0.8in]{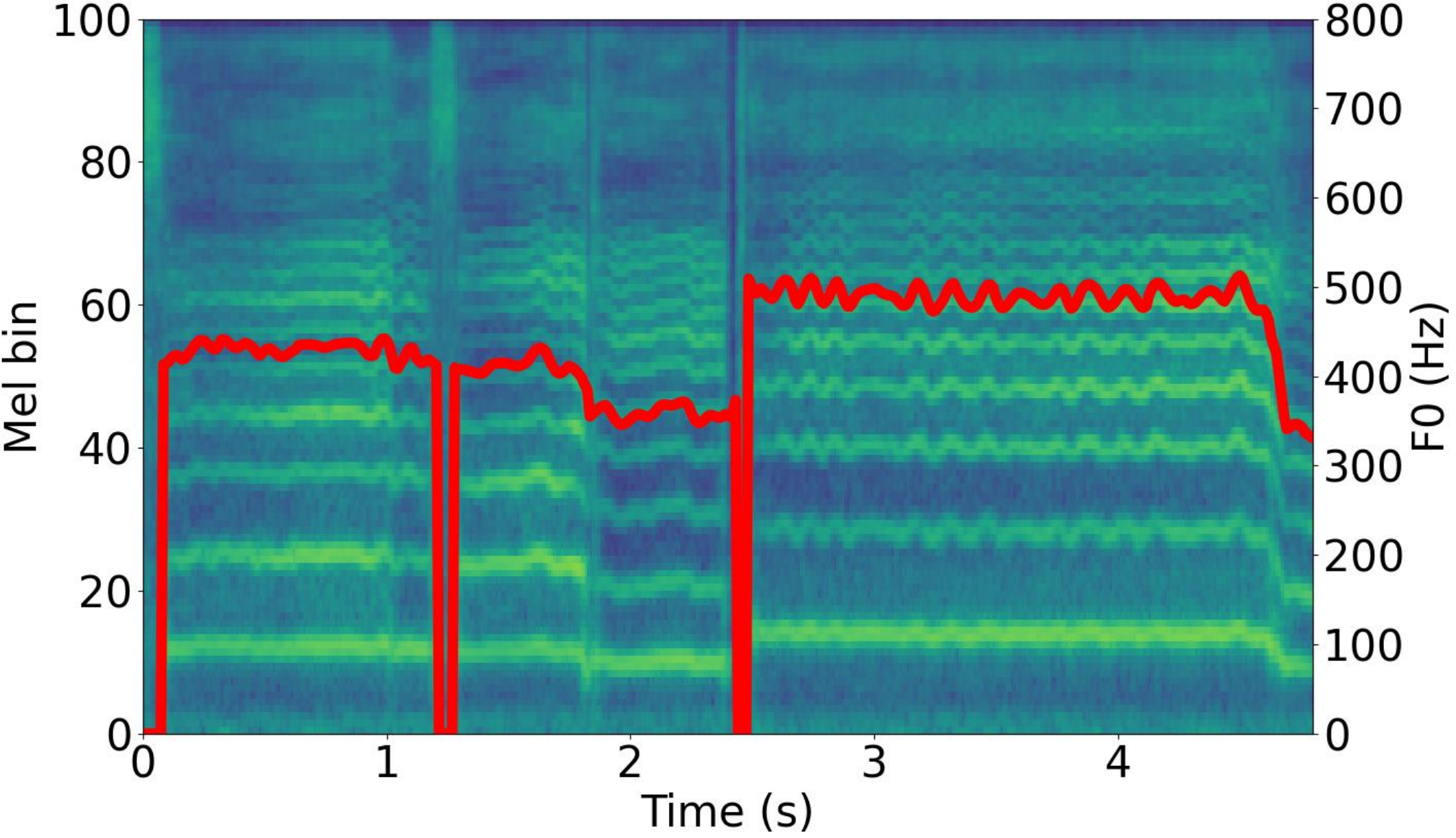}%
\label{fig:rate_control_ext1.5_rate_0.5}}
\hfil
\subfloat[$\alpha=0.5$, $\beta=1.5$]{\includegraphics[width=1.3in, height=0.8in]{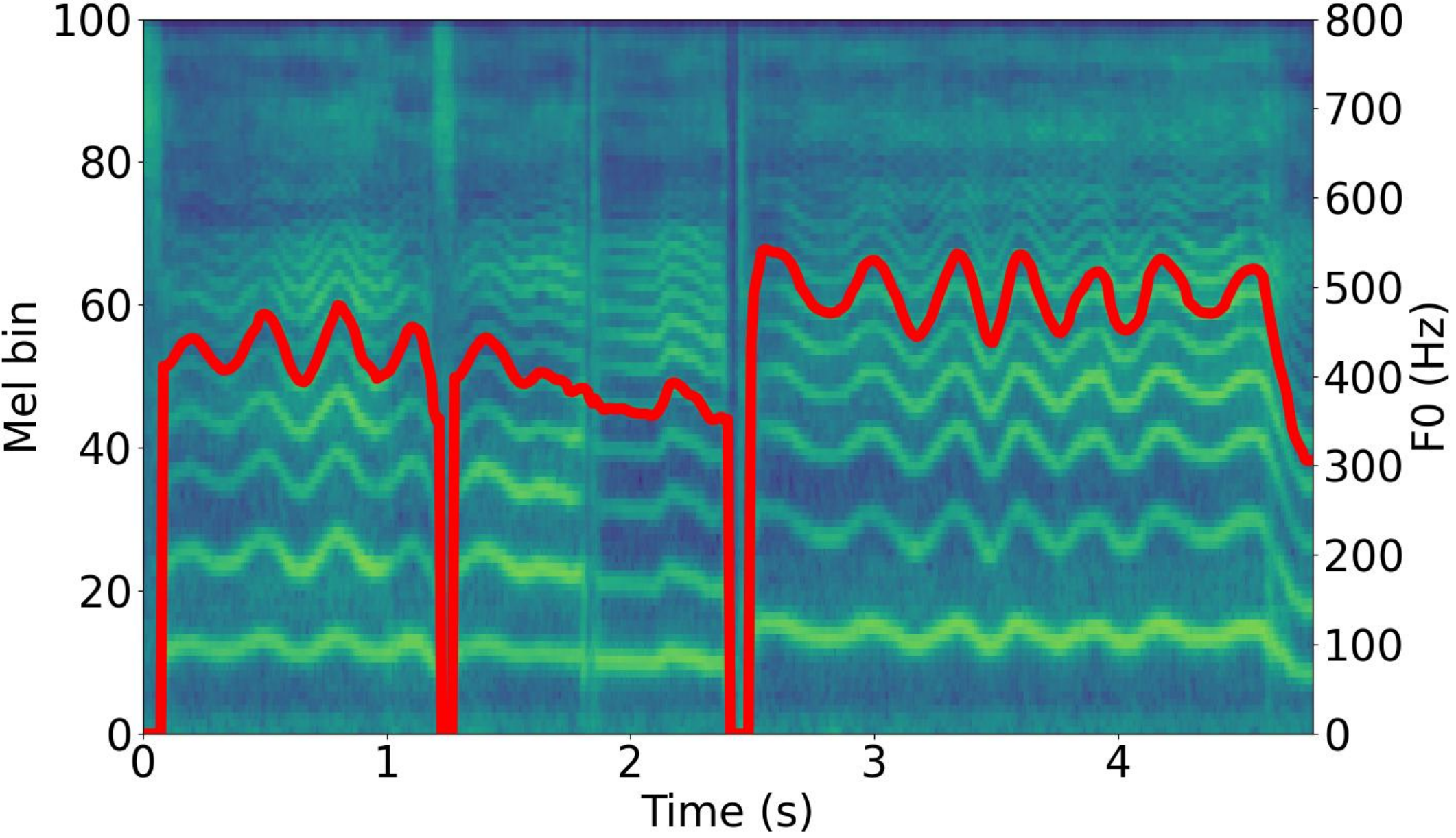}%
\label{fig:rate_control_ext0.5_rate_1.5}}
\hfil
\subfloat[$\alpha=1.5$, $\beta=1.5$]{\includegraphics[width=1.3in, height=0.8in]{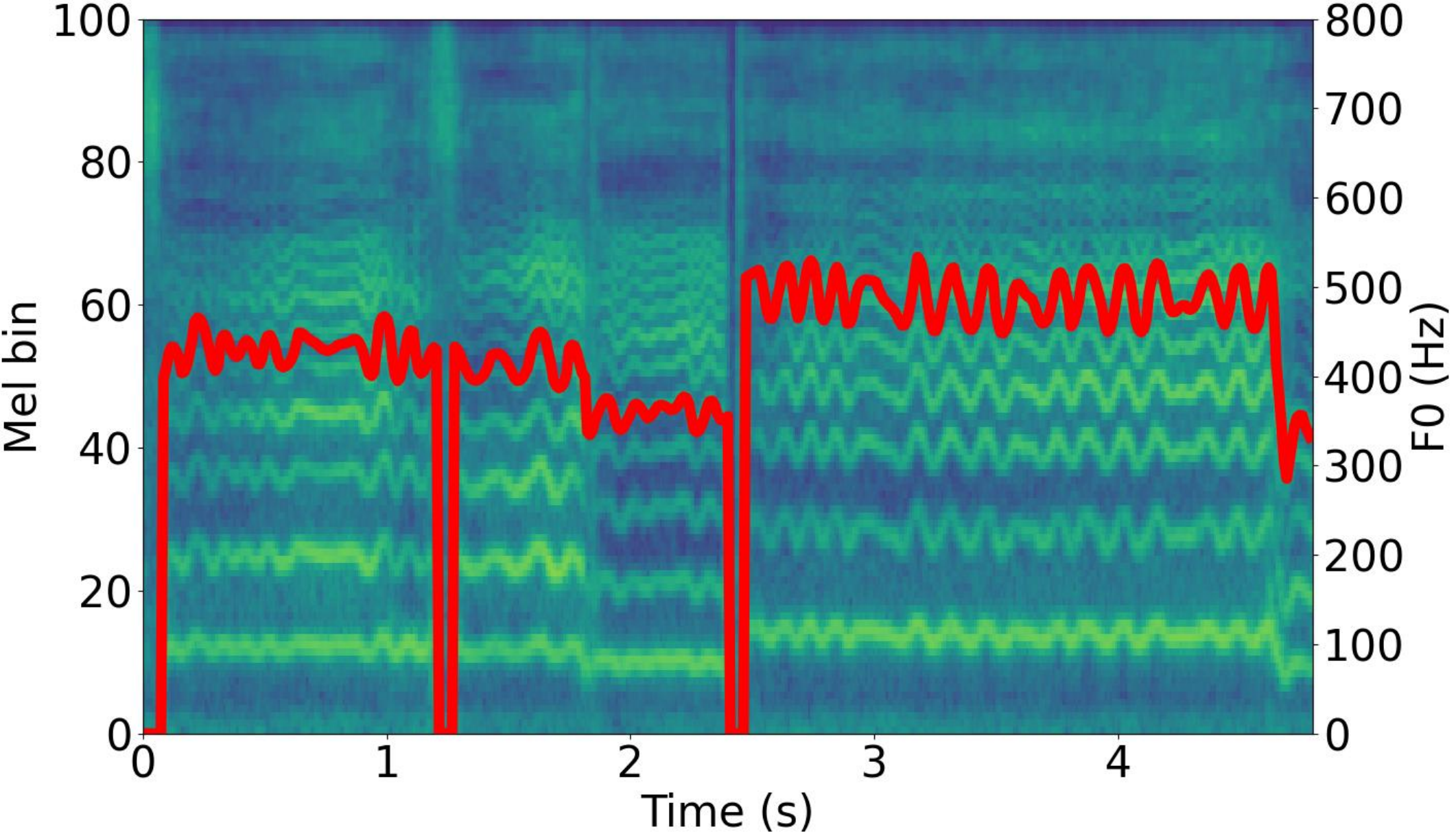}%
\label{fig:rate_control_ext1.5_rate_1.5}}
\caption{Mel-spectrograms demonstrating the independent controllability of vibrato extent ($\alpha$) and rate ($\beta$) using various scaling factors. (f) presents the result of the Straight $\xrightarrow{}$ Vibrato conversion ($\alpha=1.0$, $\beta=1.0$). (b)-(e) show the results of controlling only the vibrato rate while keeping $\alpha=1.0$. (g)-(j) show the results controlling both the vibrato rate and extent concurrently.}  
\label{fig:rate_control_mel_spec}
\vspace{-5px}
\end{figure*}
\begin{figure*}[!t]
\centering
\subfloat[F0 extent scaling]{\includegraphics[width=1.45in]{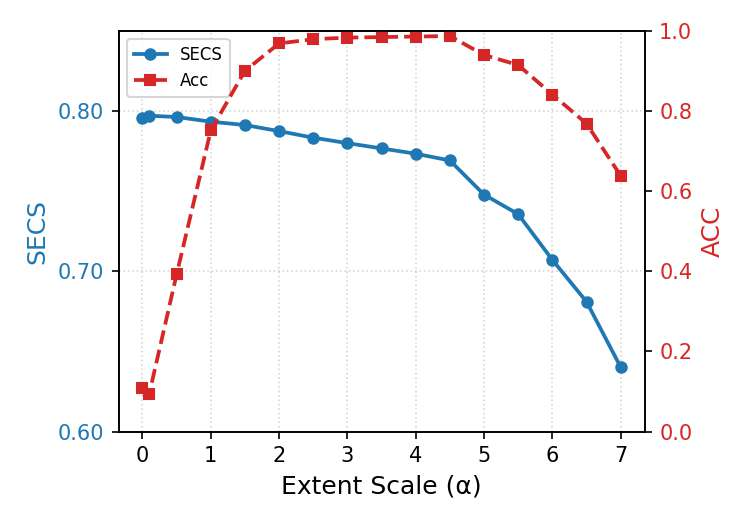}
\label{fig:scaling_boundary_extent_exp1}} 
\hspace{-0.4cm}
\subfloat[F0 rate scaling]{\includegraphics[width=1.45in]{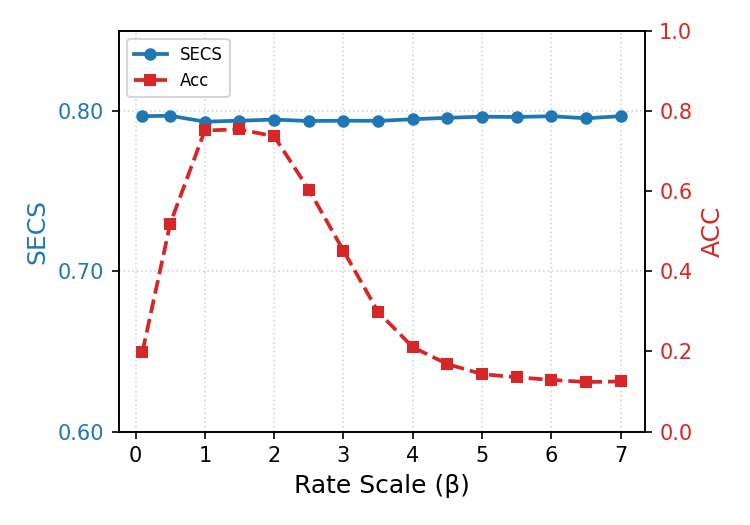}
\label{fig:scaling_boundary_rate_exp1}}
\hspace{-0.4cm}
\subfloat[Energy extent scaling]{\includegraphics[width=1.45in]{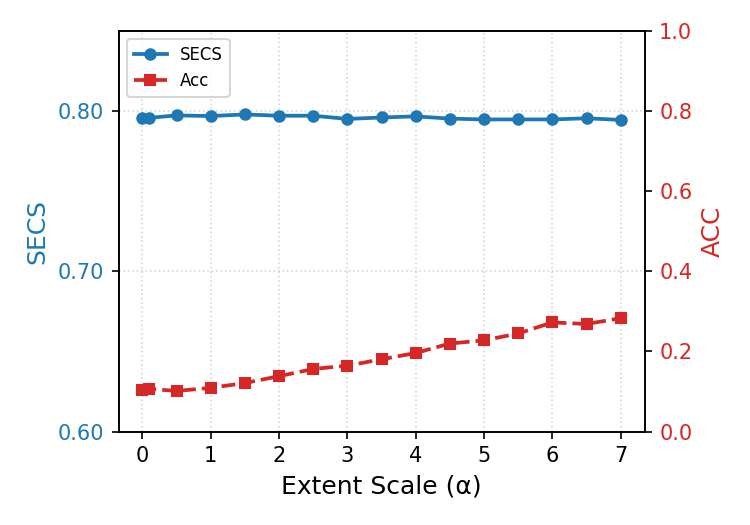}
\label{fig:scaling_boundary_extent_exp2}}
\hspace{-0.4cm}
\subfloat[Energy rate scaling]{\includegraphics[width=1.45in]{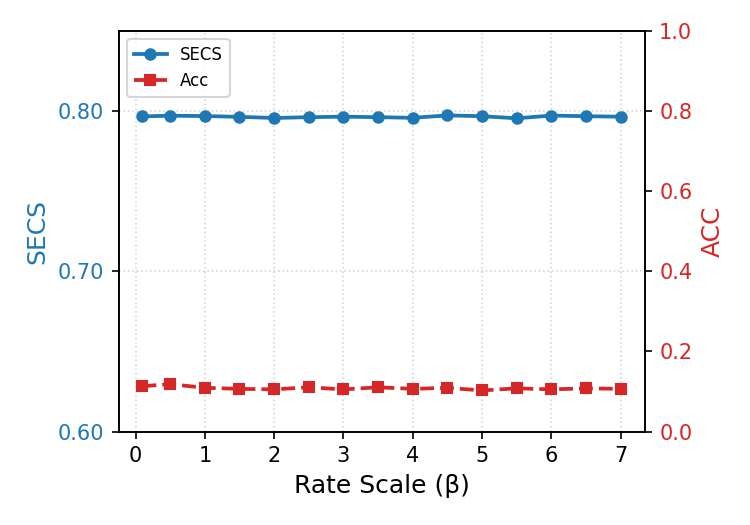}
\label{fig:scaling_boundary_rate_exp2}}
\hspace{-0.4cm}
\subfloat[Energy rate scaling ($\alpha=7.0$)]{\includegraphics[width=1.45in]{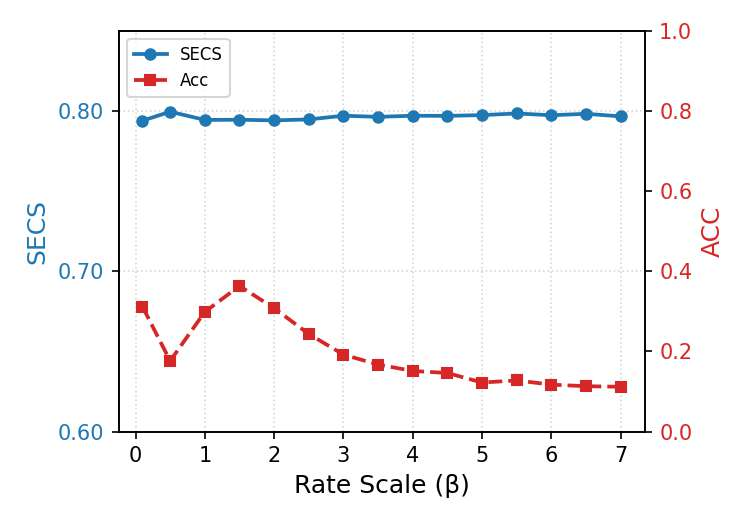}
\label{fig:scaling_boundary_rate_exp3}}

\caption{Boundary analysis of SECS (blue) and Style Accuracy (red) across various vibrato extent and rate scaling factors.}
\label{fig:scaling_boundary_exp}
\vspace{-1pt}
\end{figure*}
\begin{table}[t]
\caption{The results of the zero-shot pitch style conversion for unseen glissando style.\label{tab:glissando}}
\renewcommand{\arraystretch}{1.5}
\centering

\resizebox{1.00\linewidth}{!}{
\begin{tabular}{l||cccc}

\hline

\multicolumn{1}{l||}{\textbf{Model}}&
\multicolumn{1}{c}{\textbf{nMOS}}&
\multicolumn{1}{c}{\textbf{sMOS}}&
\multicolumn{1}{c}{\textbf{SECS}}&
\multicolumn{1}{c}{\textbf{FPC}}\\
\hline

\textbf{Seed-SVC}
& 3.921 $\pm$ 0.057     & \textbf{3.347 $\pm$ 0.042}
& \textbf{0.934}        & 0.788\\

\textbf{NeuCoSVC2}
& \textbf{3.990 $\pm$ 0.059}     & \underline{3.324 $\pm$ 0.042}
& \underline{0.918}                 & 0.790\\
\hline

\textbf{Serenade} 
& \underline{3.933 $\pm$ 0.061}     & 3.295 $\pm$ 0.043
& \underline{0.918}                 & \underline{0.794}\\
\hline

\textbf{VibE-SVC2-ZSC (Ours)}
& 3.884 $\pm$ 0.061     & 3.283 $\pm$ 0.042
& 0.912                 & \textbf{0.812}\\
\hline
\end{tabular}}
\vspace{-7pt}
\end{table}

\subsection{Performance of Zero-Shot Pitch Style Conversion}
We compared VibE-SVC2-ZSC against recent baseline models, including Zero-shot SVC (NeuCoSVC2, Seed-SVC) and SSC models (Serenade, Vevo1.5, Vevo2).
As shown in Table \ref{tab:pitch_tech_conversion_zeroshot}, VibE-SVC2-ZSC achieves the best average style accuracy, showing a large gap between the second-best model. Serenade struggles with style conversion between straight and vibrato styles, in line with the report of
the SVCC 2025 challenge \cite{svcc2025}. On the other hand, Vevo1.5 shows slightly better conversion performance in Straight $\xrightarrow{}$ Vibrato conversion with an accuracy of 0.488 and Vevo2 shows the second-best performance in  Vibrato $\xrightarrow{}$ Straight. Nevertheless, neither baseline model outperforms VibE-SVC2-ZSC in style accuracy. As shown in Fig. \ref{fig:zero_shot}, our proposed model converts to
both pitch styles better than the baseline models. Vevo1.5 and Vevo2 generate vibrato from straight samples better than Serenade. However, Vevo1.5 and Serenade struggle with removing vibrato in Vibrato $\xrightarrow{}$ Straight conversion, showing slight compression of the vibrato extent compared to the source sample and the Serenade output.

Although our model shows the most dominant style conversion performance, NeuCoSVC2 and Seed-SVC show a higher performance in both naturalness and speaker similarity. In particular, NeuCoSVC2 achieves the highest average nMOS, sMOS, and SECS. Based on several works \cite{ning2023vits, zhou2023vits}, joint training with large-scale speech and singing datasets leads to improvements in naturalness and expressiveness. Compared to
SSC models such as Serenade, Vevo1.5 and Vevo2, our model shows competitive performance in both naturalness and speaker similarity. However, the performance gaps between our model and the best baselines fall within the 95\% confidence intervals, indicating that our model remains highly competitive in both aspects. Consequently, our framework achieves the best style conversion performance with competitive perceptual quality or speaker identity.

We further evaluate the models on unseen pitch style conversion. Since the glissando style is not used during training, we adopt this style to evaluate zero-shot capabilities for unseen pitch style. However, we exclude the Vevo1.5 and Vevo2 models from this task, since they were already trained with the glissando style of the GTSinger dataset. As shown in Table \ref{tab:glissando}, our proposed model achieves the best average performance in style transfer, showing the highest FPC result. However, our model shows slightly lower performance on speaker similarity and naturalness. These lower results of speaker similarity performance indicate that our ID-based speaker timbre modeling ability may be relatively insufficient compared to baseline models like NeuCoSVC2 and Seed-SVC that utilize reference audio.

\subsection{Performance of Vibrato Controllability}
As defined in Equation \ref{equation:recon_raw_f0}, our framework allows for the simultaneous and independent control of vibrato rate and extent at inference time. Crucially, this functionality does not require any vibrato parameters or statistics as a training input. As shown in the top row of Fig. \ref{fig:rate_control_mel_spec}, our proposed rate scaling method effectively increases the vibrato rate of the converted sample. Furthermore, our proposed model achieves successful independent control of vibrato rate and extent by the comparing results between Fig. \ref{fig:rate_control_ext1.5_rate_1.5} and Fig. \ref{fig:rate_control_ext0.5_rate_1.5}.

\begin{table}[!t]
\renewcommand{\arraystretch}{1.5}
\caption{The results of the timbre style conversion task. 'Any' denotes all styles except for the target style.\label{tab:timbre_tech_conversion}}
\centering
\setlength{\tabcolsep}{3pt}
\resizebox{1.00\linewidth}{!}{
\begin{tabular}{l||cccc}

\hline  
\multicolumn{5}{c}{\textbf{Average}} \\ 
\hline

\multicolumn{1}{l||}{\textbf{Model}}&
\multicolumn{1}{c}{\textbf{nMOS}} & 
\multicolumn{1}{c}{\textbf{sMOS}} & 
\multicolumn{1}{c}{\textbf{SECS}} & 
\multicolumn{1}{c}{\textbf{Acc}}  \\ 
\hline

\textbf{GT}      
& 4.336 $\pm$ 0.033     & 3.472 $\pm$ 0.029
& 0.775                 & 0.913 \\

\textbf{Vocoded} 
& 4.320 $\pm$ 0.033     & 3.454 $\pm$ 0.029
& 0.774                 & 0.900 \\
\hline

\textbf{VibE-SVC} 
& \underline{3.752 $\pm$ 0.059}     & 2.627 $\pm$ 0.052
& \underline{0.765}                 & \underline{0.725} \\

\textbf{VibE-SVC w/ Chroma} 
& 3.568 $\pm$ 0.065     & \underline{2.654 $\pm$ 0.054}
& 0.759                 & 0.358 \\

\hline
\textbf{VibE-SVC2 (Ours)}
& \textbf{3.763 $\pm$ 0.055}    & \textbf{2.698 $\pm$ 0.052}
& \textbf{0.766}                & \textbf{0.768}    \\
\hline
\hline

\multicolumn{5}{c}{\textbf{Any} $\xrightarrow{}$ \textbf{Straight}}\\
\hline

\multicolumn{1}{l||}{\textbf{Model}}&
\multicolumn{1}{c}{\textbf{nMOS}} & 
\multicolumn{1}{c}{\textbf{sMOS}} & 
\multicolumn{1}{c}{\textbf{SECS}} & 
\multicolumn{1}{c}{\textbf{Acc}}  \\ 
\hline

\textbf{GT}      
& 4.279 $\pm$ 0.066     & 3.489 $\pm$ 0.059
& 0.775                 & 1.000 \\

\textbf{Vocoded} 
& 4.240 $\pm$ 0.063     & 3.442 $\pm$ 0.059
& 0.772                 & 0.950 \\
\hline

\textbf{VibE-SVC} 
& \textbf{3.788 $\pm$ 0.108}    & \underline{2.516 $\pm$ 0.105}
& \underline{0.773}             & \underline{0.675} \\

\textbf{VibE-SVC w/ Chroma} 
& 3.531 $\pm$ 0.134     & 2.504 $\pm$ 0.114
& 0.767                 & 0.358 \\

\hline
\textbf{VibE-SVC2 (Ours)}
& \underline{3.754 $\pm$ 0.097}     & \textbf{2.597 $\pm$ 0.103}
& \textbf{0.774}        & \textbf{0.715}    \\
\hline
\hline

\multicolumn{5}{c}{\textbf{Any} $\xrightarrow{}$ \textbf{Belt}}\\
\hline

\multicolumn{1}{l||}{\textbf{Model}}&
\multicolumn{1}{c}{\textbf{nMOS}} & 
\multicolumn{1}{c}{\textbf{sMOS}} & 
\multicolumn{1}{c}{\textbf{SECS}} & 
\multicolumn{1}{c}{\textbf{Acc}}  \\ 
\hline

\textbf{GT}      
& 4.241 $\pm$ 0.066     & 3.410 $\pm$ 0.054
& 0.781                 & 0.800 \\

\textbf{Vocoded} 
& 4.272 $\pm$ 0.068     & 3.416 $\pm$ 0.054
& 0.778                 & 0.800 \\
\hline

\textbf{VibE-SVC} 
& \underline{3.549 $\pm$ 0.126}     & \underline{2.524 $\pm$ 0.097}
& \underline{0.769}                 & \underline{0.668} \\

\textbf{VibE-SVC w/ Chroma} 
& 3.346 $\pm$ 0.140     & 2.494 $\pm$ 0.096
& 0.764                 & 0.358 \\

\hline
\textbf{VibE-SVC2 (Ours)}
& \textbf{3.556 $\pm$ 0.117}    & \textbf{2.562 $\pm$ 0.095}
& \textbf{0.770}                & \textbf{0.722}    \\
\hline
\hline

\multicolumn{5}{c}{\textbf{Any} $\xrightarrow{}$ \textbf{Breathy}}\\
\hline

\multicolumn{1}{l||}{\textbf{Model}}&
\multicolumn{1}{c}{\textbf{nMOS}} & 
\multicolumn{1}{c}{\textbf{sMOS}} & 
\multicolumn{1}{c}{\textbf{SECS}} & 
\multicolumn{1}{c}{\textbf{Acc}}  \\ 
\hline

\textbf{GT}      
& 4.274 $\pm$ 0.065     & 3.509 $\pm$ 0.060
& 0.788                 & 0.875 \\

\textbf{Vocoded} 
& 4.240 $\pm$ 0.063     & 3.491 $\pm$ 0.060
& 0.788                 & 0.875 \\
\hline

\textbf{VibE-SVC} 
& \underline{3.536 $\pm$ 0.119}     & 2.774 $\pm$ 0.105
& \textbf{0.757}                 & \underline{0.787} \\

\textbf{VibE-SVC w/ Chroma} 
& 3.480 $\pm$ 0.134     & \underline{2.808 $\pm$ 0.110}
& 0.753                 & 0.356 \\
\hline

\textbf{VibE-SVC2 (Ours)}
& \textbf{3.654 $\pm$ 0.112}    & \textbf{2.868 $\pm$ 0.106}
& \textbf{0.757}                & \textbf{0.858}    \\
\hline
\hline

\multicolumn{5}{c}{\textbf{Any} $\xrightarrow{}$ \textbf{Vocal Fry}}\\
\hline

\multicolumn{1}{l||}{\textbf{Model}}&
\multicolumn{1}{c}{\textbf{nMOS}} & 
\multicolumn{1}{c}{\textbf{sMOS}} & 
\multicolumn{1}{c}{\textbf{SECS}} & 
\multicolumn{1}{c}{\textbf{Acc}}  \\ 
\hline

\textbf{GT}      
& 4.482 $\pm$ 0.062     & 3.492 $\pm$ 0.061
& 0.756                 & 0.975 \\

\textbf{Vocoded} 
& 4.466 $\pm$ 0.062     & 3.477 $\pm$ 0.061
& 0.757                 & 0.975 \\
\hline

\textbf{VibE-SVC} 
& \textbf{4.012 $\pm$ 0.104}    & 2.719 $\pm$ 0.106
& \underline{0.762}             & \underline{0.771} \\

\textbf{VibE-SVC w/ Chroma} 
& 3.803 $\pm$ 0.106     & \textbf{2.850 $\pm$ 0.111}
& 0.753                 & 0.355 \\
\hline

\textbf{VibE-SVC2 (Ours)}
& \underline{3.984 $\pm$ 0.101}     & \underline{2.796 $\pm$ 0.106}
& \textbf{0.763}                    & \textbf{0.778}    \\
\hline
\end{tabular}}
\end{table}
\begin{table}[!t]
\caption{The results of the vocal fry enforcement.\label{tab:forcing_subharmonic}}
\renewcommand{\arraystretch}{1.3}
\centering
\setlength{\tabcolsep}{10pt}
\begin{tabular}{l|c||cc}
\hline
\textbf{Model}  &\textbf{Pitch Style} & 
\textbf{SECS}   &\textbf{Acc}\\
\hline
\textbf{VibE-SVC2} 
& \textbf{Original} & \textbf{0.763} & 0.778 \\
& \textbf{Straight} & \textbf{0.765} & 0.754 \\
& \textbf{Vibrato} & \textbf{0.758} & 0.851 \\

\hline
\textbf{VibE-SVC2}    & \textbf{Original}     
& 0.743     & \textbf{0.976} \\
\textbf{w/ Vocal Fry Enforcement}  & \textbf{Straight} 
& 0.744     & \textbf{0.977} \\
& \textbf{Vibrato} 
& 0.740     & \textbf{0.984} \\
\hline
\end{tabular}
\vspace{-3pt}
\end{table}

To evaluate these valid operational control boundaries, we conducted scaling experiments measuring both speaker similarity and style accuracy up to extreme boundaries. We synchronously scaled the vibrato extent and rate in both F0 and energy contours. First, we progressively increased the extent scaling factor $\alpha$ from 0.0 to 7.0. As shown in Fig. \ref{fig:scaling_boundary_extent_exp1}, SECS consistently decreased, while the style accuracy improved steadily up to an optimal boundary ($\alpha \approx 3.0$) before degrading. Second, we increased the rate scaling factor $\beta$ from 0.1 to 7.0, since setting it to 0.0 results in a zero-length input. SECS remained stable across all variations, as shown in Fig. \ref{fig:scaling_boundary_rate_exp1}. However, the style accuracy showed a sharp drop when $\beta$ exceeded 2.0 as visually confirmed in Fig. \ref{fig:rate_control_ext1.0_rate_2.0}. 

Although the rate scaling operation itself works effectively even at 2.0, the generated audio becomes perceptually unnatural. This aligns with music theory, as typical singing vibrato has a range of 3-9 Hz.Pushing the rate above this limit results in a mechanical vibration rather than a musical expression. Second, to validate energy contour disentanglement, we conducted an energy-only scaling experiment by fixing the F0 extent to 0.0. By scaling only the energy extent, the model generated a volume vibrato effect without any pitch variations. Although the absolute style accuracy was predictably lower, the isolated energy vibrato still provided a positive contribution to the overall style accuracy. This outcome also validates the necessity of the energy style modeling.

As shown in Fig. \ref{fig:scaling_boundary_rate_exp2}, scaling only the energy rate resulted in stable trends in SECS and style accuracy. 
To determine whether this stability was due to a lack of style information or an insufficient energy extent, we conducted an additional experiment by increasing the energy extent to 7.0, illustrated as Fig. \ref{fig:scaling_boundary_rate_exp3}. Even under this condition, energy rate scaling showed trends similar
to the F0 rate scaling results. This indicates that a sufficient extent of energy is required to control the vibrato energy rate.
However, the slightly higher style accuracy observed at 0.1 suggests a minor entanglement between the energy extent and the rate. 

\begin{figure}[!t]
\centering
\subfloat[Straight pitch style]{\includegraphics[width=1.65in]{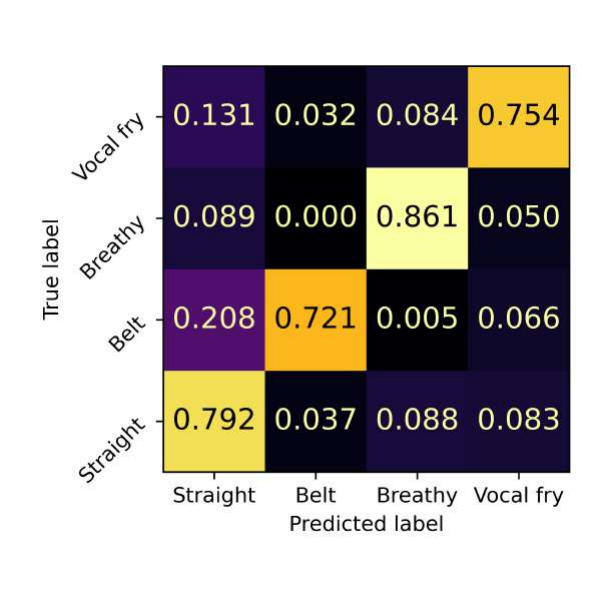}
\label{fig:confusion_straight}}
\subfloat[Vibrato pitch style]{\includegraphics[width=1.65in]{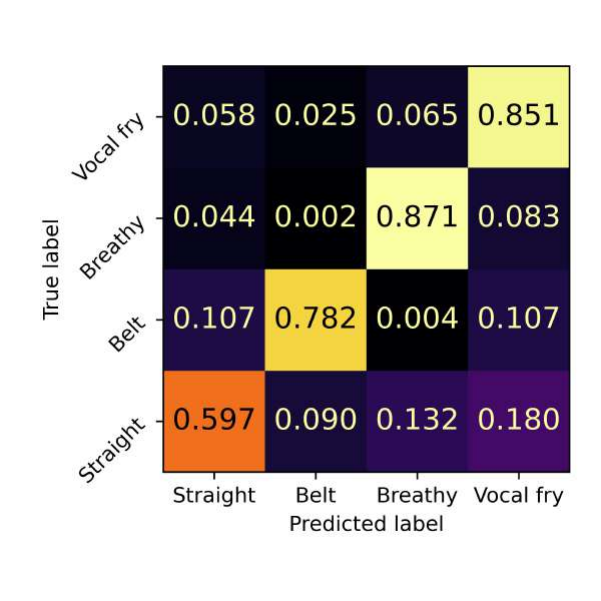}
\label{fig:confusion_vibrato}}

\caption{Timbre style accuracy results when jointly converting with different pitch styles. (a) and (b) fix the target pitch styles of style converters to straight and vibrato, respectively.
}
\label{fig:timbre_confusion}
\vspace{-5pt}
\end{figure}
\subsection{Performance of Timbre Style Conversion}
As shown in Table \ref{tab:timbre_tech_conversion}, our proposed model outperforms baseline models in terms of speaker similarity and style conversion ability in objective evaluations. 
Specifically, compared to VibE-SVC, the style accuracy to convert to
straight, belt, and breathy styles is significantly increased, respectively. However, the performance difference in 'Any $\xrightarrow{}$ Vocal Fry' conversion is marginal. These results show that our proposed SHC algorithm is effective when the source audio is vocal fry in the timbre style conversion. 
For subjective evaluations, our model outperforms the baseline models on average, showing slight degradation within the confidence interval in the vocal fry and straight conversion. 
In contrast, the baseline model, VibE-SVC with chromagram, shows some success with vocal fry conversion. However, it often suffers from pitch-jumping artifacts as a result of the lack of octave information in its input, leading to an unnatural pitch.

We also evaluated the pitch style conversion performance to assess the joint controllability of the proposed model. As shown in Fig. \ref{fig:timbre_confusion}, our proposed model enables the disentanglement and the control of these two stylistic aspects independently. However, the matrix of vibrato conversion reveals a slight degradation when the model converts the timbre style into a straight timbre while simultaneously applying a vibrato pitch. It is expected that the stability implied by a straight timbre is inherently at odds with the dynamic modulation of vibrato.

We conducted an experiment of utilizing the subharmonic characteristic to enforce the vocal fry style. To achieve this, we halved the source F0 contour as input. As shown in Table \ref{tab:forcing_subharmonic}, this forcing method results in higher style accuracy for vocal fry conversion. Specifically, scaling down the F0 range reflects the natural acoustic relationship between the extreme low F0 range and vocal fry. We speculate that low range of F0 contour guides the SVC model to add vocal fry styles, since the model is trained with halved F0 for vocal fry sample.

However, this enforcement leads to a degradation in speaker similarity. This indicates a trade-off that the F0 shift enhances the target phonation style but simultaneously deviates from the target speaker identity, which was modeled at a higher F0 range. These results demonstrate that the entanglement of the timbre style and the F0 range slightly remains.

\begin{table}[!t]
\renewcommand{\arraystretch}{1.3}
\caption{The evaluation effectiveness between energy style modeling and intelligibility.\label{tab:ablation_energy}}
\centering
\setlength{\tabcolsep}{8pt}

\begin{tabular}{l||cccc}
\hline  

\multicolumn{1}{l||}{\textbf{Model}}& 
\multicolumn{1}{c}{\textbf{SECS}} & 
\multicolumn{1}{c}{\textbf{WER}} \\ 
\hline

\textbf{VibE-SVC2-ZSC w/o ZSE}  & 0.895     & 19.80 \\
\textbf{VibE-SVC2-ZSC (Ours)}   & \textbf{0.922}     & \textbf{21.50}  \\
\hline

\end{tabular}
\vspace{3pt}
\end{table}
\begin{table}[!t]
\caption{Comparison of scaling methods in SHC.\label{tab:shc_ablation}}

\renewcommand{\arraystretch}{1.3}
\centering
\setlength{\tabcolsep}{15pt}
\begin{tabular}{l||cc}
\hline
\textbf{Scaling Method}   & 
\textbf{SECS}   &\textbf{Acc}\\
\hline
\textbf{Mean} 
 & 0.766 & 0.730 \\

\textbf{Mode}   
& \textbf{0.773}     & \textbf{0.771} \\

\hline
\end{tabular}
\end{table}

\subsection{Ablation Studies}
\subsubsection{Impact of Energy Style Modeling}
We conducted experiments to evaluate the effectiveness of the proposed energy style modeling by assessing speaker similarity and intelligibility on vibrato style conversion. Since the VocalSet dataset cannot be transcribed due to its vowel phonation, we evaluate pitch style conversion using the GTSinger dataset. Since MERT is not pre-trained on the GTSinger dataset and the pitch style classifier struggles to evaluate style accuracy, we only evaluate on SECS and WER. As shown in Table \ref{tab:ablation_energy}, adopting this module results in a marginal increment in WER. More importantly, explicitly modeling the energy dynamics improves the similarity of the target speaker.
This demonstrates that fine-grained and high-frequency energy contours have an effect on target speaker similarity.

\subsubsection{Comparison of Scaling Method in SHC}
We compared the performance of the SHC algorithm using two different scaling methods: mean alignment and mode alignment. As shown in Table \ref{tab:shc_ablation}, the mode alignment significantly outperforms the mean alignment in both speaker similarity and style conversion accuracy. Specifically, applying mean alignment resulted in a lower style accuracy compared to mode alignment. This performance degradation occurs because the mean value is sensitive to residual pitch jumps that the threshold fails to filter, shifting the entire contour toward the subharmonic range. This downward bias leads the model to erroneously enforce a vocal fry style even for non-vocal fry targets. Consequently, we confirmed that mode alignment is a more robust scaling method for subharmonic correction.

\subsubsection{Speaker-Preserving Timbre Style Conversion}
To evaluate timbre style conversion performance independent of speaker identity conversion performance, we conducted an ablation study focusing on speaker-preserving timbre transfer using the VocalSet dataset.
As detailed in Table \ref{tab:general_ssc}, the integration of the SHC algorithm improves the accuracy of the timbre style from 0.672 to 0.794 compared to VibE-SVC model. VibE-SVC2 successfully demonstrates superior disentanglement and robust control over timbre styles without heavily compromising the source speaker's inherent identity, a capability notably lacking in models like the Vevo series, which exhibits a severely diminished capacity for timbre style conversion. 
For zero-shot SVC models, the models struggle with independent control of speaker identity and timbre style. This entanglement trades off SECS for higher style accuracy.

\begin{table}[!t]
\renewcommand{\arraystretch}{1.3}
\caption{The results of speaker-preserving timbre style conversion. \label{tab:general_ssc}}
\centering

\setlength{\tabcolsep}{12pt}

\begin{tabular}{l||cc}

\hline  

\multicolumn{1}{l||}{\textbf{Model}}&
\multicolumn{1}{c}{\textbf{SECS}} & 
\multicolumn{1}{c}{\textbf{Acc}}  \\ 
\hline

\textbf{NeuCoSVC2} 
& 0.772                & 0.430 \\
\textbf{Seed-SVC} 
& 0.764                 & 0.398 \\
\hline

\textbf{Serenade} 
& 0.776                 & 0.549 \\

\textbf{Vevo1.5} 
& 0.781                 & 0.113 \\

\textbf{Vevo2} 
& \underline{0.785}                 & 0.112 \\

\textbf{VibE-SVC} 
& \textbf{0.788}                 & \underline{0.672} \\
\hline
\textbf{VibE-SVC2 (Ours)}
& 0.779        & \textbf{0.794}    \\
\hline
\end{tabular}
\end{table}

\subsection{Analysis of Trade-offs between Metrics}
\subsubsection{Style Accuracy and Speaker Similarity}
Manipulating F0 beyond natural boundaries causes a trade-off between style intensity and speaker identity. As shown in Table \ref{tab:forcing_subharmonic} and Fig. \ref{fig:scaling_boundary_extent_exp1}, lowering F0 or increasing F0 extent decreases SECS. This occurs because extreme F0 range shifts push acoustic features outside the target speaker's distribution. Therefore, increasing style intensity reduces speaker similarity, requiring a balance between style expressiveness and identity preservation depending on the application's priorities.

\subsubsection{Style Controllability and Naturalness}
An architectural trade-off exists between naturalness and style controllability, as shown in Table \ref{tab:pitch_tech_conversion_zeroshot} and Table \ref{tab:glissando}. While structural decoupling enables independent style transfer, it degrades naturalness compared to fully entangled baselines. This happens because the ID-based lookup table filters out acoustic nuances from the reference audio. Thus, ID-based architecture should be applied selectively, depending on whether the application requires precise timbre and timbre style manipulation or maximum naturalness.

\section{Conclusion}
In this paper, we propose a singing voice conversion model that improves the singing style conversion performance and expands the singing style controllability. 
We improve the pitch style conversion performance by considering the periodic information of the pitch style using the energy style converter. In addition, we propose ZSC module, which converts the pitch style to match that of a reference audio. We further introduce vibrato rate control, enabling independent control of vibrato rate and extent. We expand the model to include joint control of the timbre style and propose the SHC algorithm to improve the timbre style conversion performance. The algorithm successfully improves the timbre style conversion performance without degradation in terms of naturalness and speaker similarity. We demonstrate the effectiveness of our proposed methods through comprehensive objective and subjective evaluations.

\section{Future Works}
We plan to improve and extend our framework in three directions. First, we are planning to expand the model into a unified model with SVS and music generation. In addition, we will try to expand the model to control singing styles through a natural language prompt. 
Second, we will expand VibE-SVC2 with the ZSC module into a fully zero-shot singing style conversion model, capable of transferring timbre styles from an arbitrary reference audio. A primary focus will be on achieving a more robust disentanglement of the timbre style from the speaker identity. 
Lastly, we will address a current limitation in our proposed SHC algorithm. Due to its autoregressive nature, this can lead to an accumulation of errors, particularly when an audio sample contains many unvoiced regions. Although this issue does not significantly degrade timbre conversion performance, it can negatively impact melody similarity. We aim to improve the robustness of the algorithm by incorporating mechanisms to handle these unvoiced segments more effectively, thereby enhancing the overall pitch fidelity of the conversion.


\bibliographystyle{IEEEtran}
\bibliography{bibtex}
\vspace{-20pt}

\begin{IEEEbiography}[{\includegraphics[width=1in,height=1.25in,clip,keepaspectratio]{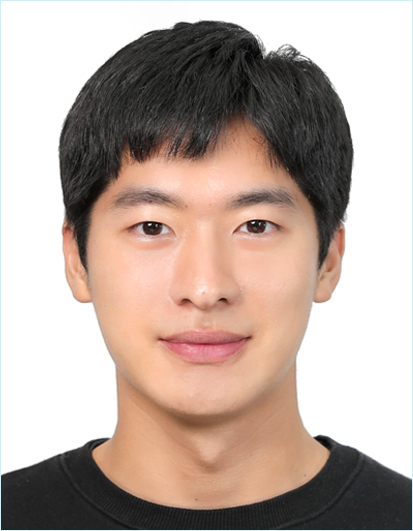}}]{Joon-Seung Choi} received the B.S. degree in computer science from the Inha University, Incheon,
South Korea, in 2024. He is currently pursuing
toward the master’s degree with the Department of Artificial Intelligence, Korea University, Seoul, South Korea. His research interests include artificial intelligence and audio signal processing.
\end{IEEEbiography}

\begin{IEEEbiography}[{\includegraphics[width=1in,height=1.25in,clip,keepaspectratio]{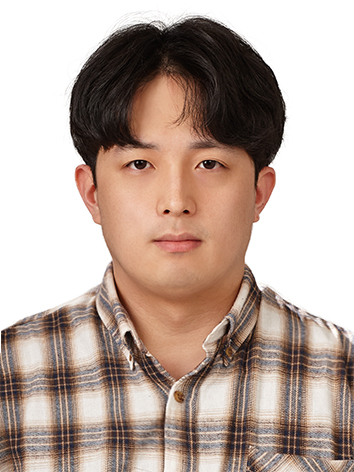}}]{Dong-Min Byun} received the B.S. degree in radiological science from the Yonsei University, Wonju,
South Korea, in 2021. He is currently working toward
the master’s and Ph.D. degrees with the Department
of Artificial Intelligence, Korea University, Seoul,
South Korea. His research interests include artificial
intelligence and audio signal processing.
\end{IEEEbiography}

\begin{IEEEbiography}[{\includegraphics[width=1in,height=1.25in,clip,keepaspectratio]{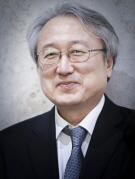}}]{Seong-Whan Lee} (Fellow, IEEE) received the B.S. degree in computer science and statistics from Seoul National University, Seoul, Republic of Korea, in 1984, and the M.S. and Ph.D. degrees in computer science from the Korea Advanced Institute of Science and Technology, Seoul, Republic of Korea, in 1986 and 1989, respectively. He is currently the Head of the Department of Artificial Intelligence, Korea University, Seoul, Republic of Korea. His current research interests include artificial intelligence, pattern recognition, and brain engineering. He is a fellow of the International Association of Pattern Recognition (IAPR), the Korea Academy of Science and Technology, and the National Academy of Engineering of Korea.
\end{IEEEbiography}

\vfill

\end{document}